\documentclass[12pt,preprint]{aastex}
\newcommand{\be}{\begin{equation}}
\newcommand{\ee}{\end{equation}}
\newcommand{\lmf}{\overline{\log M}_f}
\newcommand{\ltorder}{\hbox{ \rlap{\raise 0.425ex\hbox{$<$}}\lower
0.65ex\hbox{$\sim$} }}
\newcommand{\gtorder}{\hbox{ \rlap{\raise 0.425ex\hbox{$>$}}\lower
0.65ex\hbox{$\sim$} }}
\shorttitle{M87 globular cluster system}
\shortauthors{Vesperini et al.}
\begin{document}
\title {Modeling the dynamical evolution of the M87 globular cluster system}
\author{E.Vesperini\altaffilmark{1}, S.E. Zepf\altaffilmark{1}, A. Kundu\altaffilmark{1}}
\affil{Department of Physics and Astronomy, Michigan State University, East Lansing, MI, 48824 USA}
\altaffiltext{1}{e-mail:vesperin, zepf, akundu @pa.msu.edu}
\author{K.M. Ashman\altaffilmark{2}}
\affil{Department of Physics, University of Missouri, Kansas City, MO,
64110 USA} 
\altaffiltext{2}{e-mail: ashmank@umkc.edu}
\begin{abstract}
We study the dynamical evolution
of the M87 globular cluster system (GCS) with a number of numerical
simulations. We explore a range of different initial
conditions for the GCS mass function (GCMF), for
the GCS spatial distribution and for the GCS velocity distribution.
Our simulations include the effects of two-body relaxation, dynamical
friction and mass loss due to stellar evolution. 

We first confirm that
an initial power-law GCMF like that observed in young cluster systems
can be readily transformed through dynamical processes into a bell-shaped
GCMF. However, only models with 
initial velocity distributions characterized by a strong radial
anisotropy increasing with the galactocentric distance are able
to reproduce the observed constancy of the GCMF at all radii.
We show that such strongly radial orbital distributions are
inconsistent with the observed kinematics of the M87 globular
cluster system. 

The evolution of models with a bell-shaped
GCMF with a turnover similar to that currently observed in old GCS
is also investigated. We show that models with this initial GCMF can
satisfy all the observational constraints currently available on the
GCS spatial distribution, the GCS velocity distribution and on the
GCMF properties. In particular these models successfully reproduce 
both the lack of a radial gradient of the GCS mean mass recently found
in an analysis of HST images of M87 at multiple locations, and
the observed kinematics of the M87 GCS. 
Our simulations also show that evolutionary processes significantly
affect the initial GCS properties by leading to the disruption of many
clusters and changing the masses of those which survive. 
The preferential disruption of inner clusters flattens the initial GCS
number density profile and it can explain the rising specific frequency
with radius; we show that the inner flattening
observed in the M87 GCS spatial distribution can be the result of the 
effects of dynamical evolution on an initially  steep density profile. 

\end{abstract}
\section{Introduction}
A number of recent observational studies have provided 
detailed information on the properties of the M87 GCS.
Most recently, Kundu, Zepf \& Ashman (2002)
have used multiple HST pointings to show that  the 
turnover  of the luminosity function of the GCS (hereafter we will
indicate the luminosity function and the mass function of 
the globular cluster system by GCLF and GCMF respectively) is constant
over a large range of galactocentric distances.
The data analyzed include clusters up to a projected
galactocentric distance equal to about 70 kpc and
confirm and extend the lack of radial gradient in the GCLF turnover
already found by Kundu et  
al. (1999) for clusters in the inner regions of M87.
Because of the very large number of clusters in the M87 GCS and the
high quality of the HST data, the
uncertainty in the determination of the GCLF  turnover is small
and a  stringent constraint on its possible 
radial variation is thus imposed by the new data.

The lack of a radial gradient in
the GCS mean luminosity is sometimes  interpreted as an indication that 
evolutionary processes, whose efficiency is known to depend on the
galactocentric distance, did not play an important role in determining
the current properties of GCSs. 
However, the studies of the dynamical processes and their effects  
on the properties of globular clusters are based on fundamental
stellar dynamics and there is no way to keep
dynamical evolution from occurring. The 
results of all the theoretical investigations of 
the dynamical evolution of globular clusters and GCSs (see e.g. Fall
\& Rees 1977,   
Gnedin \& Ostriker 1997, Murali \& Weinberg 1997a,b, Baumgardt 1998,
Vesperini 1997, 1998, 2000, 2001 and references therein) show that
evolutionary processes lead 
to the disruption of a significant number of clusters (the exact
fraction of the initial number of clusters depending on the properties
of the host galaxy; see e.g. Murali \& Weinberg 1997a, Vesperini 2000,
2001) and to the modification of the properties of the surviving
clusters.

Several recent papers on models of the dynamical evolution of GCSs
have addressed the issue of the radial behavior of the GCMF.
Vesperini (1998, 2000), in studies of the  evolution
of the Galactic GCS and of GCSs of elliptical galaxies, has shown that
for a log-normal initial  
GCMF (or more  in general for GCMFs which are bell-shaped in
$\log M$; see Vesperini 2002)
with mean mass and dispersion similar to those currently observed,
evolutionary  
processes are very efficient in disrupting a large fraction of
clusters but they do not always  produce a significant radial variation of
the GCMF parameters. 
On the other hand, several theoretical
investigations (see e.g. Baumgardt 1998, Murali \& Weinberg 1997a,b,
Vesperini 1998,2001) have shown that for a
power-law initial GCMF as steep as 
that observed in young GCS in merging galaxies, evolutionary processes 
lead to the development of mean mass radial gradients not consistent with 
observations. 

In a recent study of the evolution of the Galactic GCS, Fall \& Zhang
(2001) have shown that, adopting  an initial orbital distribution
characterized by a strong radial anisotropy increasing with the
galactocentric distance, the final GCMF has a modest radial variation
that is in fair agreement with the observed one, even for a
steep power-law for the initial GCMF. 
Nevertheless, the radial anisotropy necessary to obtain the
desired results is very large and, although the initial anisotropy can
be reduced by disruption of clusters on more eccentric orbits,
observational estimates of the current anisotropy of the Galactic GCS
(see e.g Dinescu, Girard \& van Altena 1999) indicate a much smaller
anisotropy.  

In this paper we will model the evolution of the M87 GCS and explore
the range of initial conditions leading to final properties consistent
with those  observed. The GCLF properties determined by Kundu et
al. (2002) and the kinematical properties determined by Cote et
al. (2001), Romanowsky \& Kochanek (2001), Cohen (2000) and
Cohen \& Ryzov (1997)
will be used to constrain viable initial conditions.  
The outline of this paper is the following.
In section 2 we describe the method used for our study and the initial
conditions considered; in section 3 we show the results obtained, in
section 4 we discuss our results  and in
section 5 we summarize our conclusions.

\section{Method and initial conditions}

 The method we use to follow the evolution of a model GCS is the same as that
adopted in Vesperini (2000,2001) to study the evolution of GCS in
 elliptical galaxies. The evolution of the masses of individual
 clusters is calculated using the results of N-body simulations by
 Vesperini \& Heggie (1997). The effects of mass loss
 due to stellar evolution (assuming a stellar IMF with a slope
 $\alpha=2.3$; similar results are obtained if the 2-slope power-law
stellar IMF  suggested by Kroupa (2001) is adopted), two-body
relaxation, dynamical friction and   
 of the tidal field of the host galaxy have been included. 
The mass loss rate due to two-body relaxation has been calculated
assuming the tidal radius of each cluster is determined by the
pericentric distance. 
We refer the reader to Vesperini (2000) for further details.

We adopt a mass model for M87  proposed by McLaughlin (1999a) on
 the basis of various observational constraints available on the
 stellar and dark matter distribution in M87 and the Virgo Cluster;
 in this model, adopting a distance of 15 Mpc to M87, the total mass
 enclosed within a radius $r$ is equal to   
\begin{eqnarray}
M_{tot}(r)&=&8.1\times 10^{11}
M_{\odot}\left[(r/5.1
\hbox{kpc})/(1+r/5.1\hbox{kpc})\right]^{1.67}+ \nonumber \\ & &7.06\times 10^{14}
M_{\odot}[\ln(1+r/560\hbox{kpc})-(r/560
\hbox{kpc})/(1+r/560\hbox{kpc})].
\end{eqnarray}
For the initial spatial distribution of the GCS we have
considered the following profiles:
\begin{itemize}
\item[A)] a Navarro, Frenk \& White (1996) (hereafter NFW) profile with scale
radius, $r_s$, equal to 9.1 kpc. This model has been shown (McLaughlin
1999b) to fit the observed surface number density profile of the M87 GCS.
\item[B)] An NFW profile with $r_s=9.1$ kpc for $r<95$ kpc and an NFW profile
with $r_s=560$ kpc for $r>95$ kpc. The choice of the NFW profile with
$r_s=560$ kpc for $r>95$ kpc, where there are no observational
constraints on the GCS spatial distribution, has been proposed by
McLaughlin (1999b) and it relies on the assumption that the density
profile of the  GCS becomes the same as that of
gas and stars at $r \approx 95$ kpc.
\item[C)-D)] It has been often claimed that the inner flattening in the spatial
distribution of the M87 GCS can not  be the result of the effects of
dynamical disruption of clusters (Lauer \& Kormendy 1986).
In order to further explore if this is indeed the case we have also considered
two profiles equal to those considered in
A) and B) above but with 
$r_s=0.1$ kpc instead of the current best fit value of 9.1 kpc, giving
much more concentrated initial GCS profile. 
\end{itemize}

As for the initial GCMF, we have restricted our choice to the
functional forms and the parameters indicated by observational studies
of young and old cluster systems: our fiducial choice was a power-law
function, $f(M) 
\sim M^{-\alpha}$ for $10^4 M_{\odot}<M<10^7 M_{\odot}$ , with
index $\alpha=1.8$ as observed in young clusters systems
in merging galaxies
(see e.g. Zepf et al. 1999 and Whitmore 1999 for a review). 
Since, with an initial power-law GCMF, none of our models satisfied
all the observational constraints, we have explored the evolution of
GCSs with a 
two-slope power-law with 
$\alpha=1.8$ for $10^{5.25}~M_{\odot}<M<10^7 M_{\odot}$ and  $\alpha=0.2$ for
$10^4 M_{\odot}<M<10^{5.25}~M_{\odot}$. The latter GCMF is bell-shaped
 with a 
turnover at $\log M=5.25$ if binned in $\log M$ (see e.g. McLaughlin
1994) and is similar to 
that currently observed in old cluster systems (see e.g. Ashman \&
Zepf 1998, Harris 2001).
As discussed in Vesperini (2002), the evolution of this GCMF is not
significantly different from  that of other bell-shaped GCMFs, such as a
log-normal or a t-Student distribution, with the same turnover.

The initial density profiles and GCMFs considered are summarized in Table 1.

As for the GCS velocity distribution, we
adopted an initial distribution  
characterized by a radial anisotropy increasing with the
galactocentric distance as in the Osipkov-Merrit models (Osipkov 1979,
Merritt 1985); for these models
the ratio of the radial to the tangential velocity dispersion
increases with galactocentric distance, $r$, as

\be
{\sigma^2_R \over \sigma^2_T}=1+r^2/r_a^2. \label{eq2}
\ee
Beyond the anisotropy radius, $r_a$, the distribution is dominated by
radial orbits whereas for $r<r_a$, it is essentially isotropic.
According to the calculations of Fall \& Zhang (2001), for the Galactic
GCS, this distribution along with a 
particular choice of the anisotropy radius ($r_a=5$ kpc) appears to be
the only one leading to a GCS mean mass radial gradient in fair agreement with
observations if a power-law initial GCMF as steep as that of young
cluster systems in merging galaxies is adopted. 

In order to fully explore the dependence of the GCS evolution on the
initial anisotropy in the velocity distribution, for each of the
initial density profiles mentioned above, simulations with 
initial values of $r_a$ ranging from 1 kpc to 250 kpc have been
carried out.

Each GCS investigated was initially made of 400000 clusters with
positions randomly drawn from the spatial profiles described above;
given the current position of a cluster, $r$, the local radial velocity
dispersion was calculated using the Jeans equation and the local
tangential velocity dispersion was then calculated using the value of
${\sigma^2_R
/\sigma^2_T}$ at $r$; the radial and the tangential velocities of each 
cluster were then drawn from 
Maxwellian velocity distributions with  radial and tangential
dispersions calculated above.
This method, introduced by Hernquist (1993),
has been used in many studies to set initial conditions for
N-body simulations of both spherical and disk galaxies.

The evolution of each GCS has been followed until $t=15$ Gyr and the
GCS properties at several intermediate ages have also been studied.

\section{Results}
\subsection{Power-law initial GCMF}
In Fig. 1, we plot the mean mass of the globular
clusters against the projected galactocentric distance for theoretical
models with  
different initial anisotropy radii. We also plot the
 observational data from Kundu et al. (2002) analysis.
The absence of any significant radial gradient in the mean mass
observed for the M87 GCS requires small values of the initial
anisotropy radius ($r_a \ltorder 3$ kpc). Initial conditions with anisotropy 
radii larger than 3 kpc produce final mean masses smaller than 
those observed and steep radial gradients inconsistent with data.

We have also considered possible radial variations in $M/L_V$
for the M87 GCS. The only observed property of the globular
clusters that varies with galactocentric radius which is relevant 
for $M/L_V$ is the ratio of blue to red
globular clusters, which increases with galactocentric distance for
M87 (Kundu et al.\ 2002) as it does for most other galaxies
(see Ashman \& Zepf 1998). This color difference tends to
produce a difference in $M/L_V$ for typical models of stellar
populations, but this difference is much smaller than the
mass variation predicted by models of dynamical evolution
with a power-law initial GCMF and anisotropy radii larger
than a few kpc. Specifically, using current stellar populations
models (Bruzual \& Charlot 2000, Maraston et al. 2002) 
we find that $(M/L_V)_{red} \sim 1.6 (M/L_V)_{blue}$ 
for a range of possible age and metallicity combinations
that are consistent with the observed colors of the red and blue 
populations of globular clusters. Applying this estimate
to the M87 system, for which we approximate the color gradient
by adopting a ratio of the number of red
to blue clusters equal to 1 for $R_g<20$ kpc and equal to 0 beyond 20
kpc, we find the inner clusters would have a mass-to-light ratio
1.3 times larger than the mass-to-light ratio of outer
clusters leading  to a $\Delta \lmf \sim 0.1$. This is significantly
smaller than the radial gradient obtained in models with large initial
anisotropy radii. Moreover, the mean mass of outer clusters from
simulations  with large anisotropy radii is much smaller than the
mass obtained from the observed mean luminosity with any reasonable
value of the mass-to-light ratio.

This comparison with the observational constraints on the radial
variation of the mean mass of clusters has  significantly
narrowed the range of possible initial values of $r_a$ for the M87 GCS with a
power-law initial GCMF. 

We now turn our attention to the observational
constraints imposed by the kinematics of the M87 GCS. In fig.2 we show the 
observed projected velocity dispersion profile of the M87 GCS as
determined by Cote et  
al. (2001) along with the dispersion profiles of various models with
different orbital anisotropy. Density profile A
(fig.2a), density profile B (fig.2b) and  an
Osipkov-Merritt radial profile for the anisotropy (see eq.\ref{eq2}
above) are used to calculate the theoretical profiles.
Values of $r_a$ larger than 50 kpc  are necessary to
fit the projected velocity dispersion profile if density profile A is
adopted and larger than 200 kpc for density profile B. These values
are much larger than the initial anisotropy radius required to
produce a radial gradient of the mean mass of clusters consistent with
observations. Fig. 3 shows the final projected velocity dispersion
profiles from our simulations for initial values of $r_a$ leading to
final GCMF properties consistent with observations: it is clear that
these profiles do not fit the observational data. For the small
initial values of $r_a$ necessary to produce GCMF properties consistent with
observations, the number density profile does not evolve
significantly (see discussion below) 
and, therefore, simulations starting with profiles C and D lead to both 
kinematical properties and  final
density profiles  inconsistent with observations. 

Fig.4 shows the radial profile of $\sigma_R^2/\sigma_T^2$ at the
end of the simulations with initial 
$r_a=2$ and $r_a=3$ kpc and with initial density profile A,  and with initial
$r_a=1$ and $r_a=2$ kpc with initial density profile B:
the preferential disruption of clusters on
high eccentricity orbits is not sufficient to reduce the strong
initial anisotropy to a level consistent with observations; the
$\sigma_R^2/\sigma_T^2$ profile from the most and the least
anisotropic models used to fit the
observed velocity dispersion profile in fig. 2 along with the value of
$\sigma_R^2/\sigma_T^2$ of the different models proposed by Romanowsky \&
Kochanek (2001) calculated at the outermost distance constrained
by observations are also shown in fig.4. 

It is thus clear from these results that, adopting a  strong initial radial
anisotropy increasing with galactocentric distance, although necessary
to produce final GCMF properties 
consistent with observations from a steep initial power-law GCMF,
is not a viable solution as  the final anisotropy is stronger than
shown by observations of the radial velocities of the globular
clusters. It is also important to point out that,
although numerical simulations of galaxy formation predict a radial
anisotropy increasing with galactocentric distance, the initial
anisotropy required to obtain GCMF properties consistent with data is
much stronger than that found in simulations. For example, numerical
simulations  of the formation of giant galaxies at the 
center of groups and clusters carried out by Dubinski (1998) predict a
$\sigma_R^2/\sigma_T^2$ growing
from 0 to 2 from the center of the galaxy to 100 kpc; the initial
anisotropy required to fit the GCMF properties in our models is such
that $\sigma_R^2/\sigma_T^2\simeq 10^3$ at 100 kpc.

Fig. 5 shows the time evolution of $\lmf (R_g)$ for two models with $r_a$ 
equal to 2 and 150 kpc. In the model with initial anisotropy radius
equal to 2 kpc the mean mass profile is always approximately flat and,
as a result of the preferential disruption of low-mass clusters, the
GCS mean mass increases with time.
In the model with
$r_a=150$ kpc, a strong radial gradient quickly develops and it
becomes steeper and more extended as the system evolves; 
the quick formation of the mean mass radial gradient implies that a
significant radial gradient  would 
have to be expected in models with large anisotropy radii 
even if the time needed for the potential
of M87 to reach its equilibrium state were considered and the GCS evolved
in the current M87 potential for a time shorter than that (15 Gyr)
considered in our simulations.

As shown above, none of the initial conditions considered 
leads to final properties consistent with both the observed GCLF
properties and the observed kinematical properties. Although
it is therefore not necessary to explore the evolution of the GCS number
density profile to further constrain the range of 
viable initial conditions, we conclude this subsection with a few
remarks on the evolution 
of the GCS spatial distribution as this can shed further light on the
GCS evolution.
 
Fig.6a and 6b  show the final GCS surface density for a few different
initial values of $r_a$ along with the initial  
profile (profile A for fig.6a and profile C for fig.6b).
The
initial velocity  distribution adopted is such that the 
anisotropy increases with galactocentric distance and, in particular,
radial orbits dominate beyond $r_a$. For small values of $r_a$, most
clusters, regardless of  their current galactocentric distances,
have small pericentric distances (see fig.8a) and, since we have  
assumed that the pericentric distance determines the timescale of
evolution of clusters, the
disruption rate is approximately constant and independent of the
current galactocentric
distances of clusters (see fig.7). 
As a consequence of that, the shape of the final spatial profile is
very similar to the initial one;  this is also the reason why small values of
$r_a$ do not produce a gradient in the mean mass:
for small values of $r_a$, clusters at any distance from the galactic
center are disrupted and lose
mass at  approximately the same rate. More in general,
no radial gradient in any other cluster property will be produced as
a result of evolution for GCS with initial small values of $r_a$.
Note that if the disruption
rate does not depend on galactocentric distance, the decrease of
specific frequency in the inner 
regions of M87 reported by McLaughlin (1999b; see Rhode \& Zepf 2001
for a similar radial dependence of the specific frequency in NGC 4472)
can not be ascribed 
to dynamical evolution and it should be explained in terms of  a radial
dependency of cluster formation efficiency.

As $r_a$
increases and the velocity distribution becomes more isotropic, the
range of pericentric distances broadens (see fig.8b) and 
clusters closer to the center of the host galaxy are preferentially
disrupted (see fig.7); the disruption rate is therefore
higher near the center of the galaxy and a central flattening of the 
density profile ensues. 
Therefore we find that 
evolutionary processes, in particular disruption of inner low-mass
clusters, can be responsible 
for the observed central flattening in the spatial distribution of the
M87 GCS (see also Murali \& Weinberg 1997a, Capuzzo Dolcetta \&
Tesseri 1997) and, contrary to previous claims, this does not have to 
due to initial conditions.

\subsection{Two-slope power-law initial GCMF}
As discussed in the previous subsection, starting
with a power-law initial GCMF similar to that of young GCSs observed
in merging galaxies, we could not find any model satisfying all the
observational constraints. We have therefore continued our investigation by
exploring the evolution of GCSs with a two-slope power-law initial
GCMF. The parameters of the initial GCMF 
adopted are summarized in Table 1. We note that this GCMF is
bell-shaped 
if binned in $\log M$ and, as discussed in Vesperini (2002), the results
obtained using this functional form for the initial GCMF do not differ
significantly from those 
obtained adopting other bell-shaped
functions often used in the literature to fit the GCLF of old GCS,
such as a log-normal, or a t-Student distribution (see e.g. Secker 1992). 
As shown in the previous subsection, an initial
number density profile similar to that observed preserves its shape only for
an initial velocity distribution characterized by a strong radial
anisotropy inconsistent with 
the kinematical data available. For this reason, here, we 
focus our attention only on number density profiles C and D.

Fig.9 shows the $\lmf$ radial profile  for our simulations
of the M87 GCS with different initial values of $r_a$: the GCS final
$\lmf$ and its radial profile are consistent with the 
observed mean mass and lack of a radial gradient in the mean mass
for the entire range of values of $r_a$
considered and no constraint on the anisotropy of the initial velocity
distribution is imposed by the final GCMF properties. 

In fig.10 we compare the final projected velocity dispersion radial profile
from our simulations of M87 GCS with that determined by Cote et al's
observational data. Fig. 10a shows the results for initial number
density profile C 
and $r_a=30,~60,~200$ kpc; initial values of the anisotropy radius
around 60 kpc ($50<r_a<100$ kpc) appear to be those which better fit
the observed 
velocity dispersion profile, but all the profiles resulting from
simulations with initial
anisotropy radii ranging from 30 kpc to 200 kpc fall within the
$90\%$ confidence limits and increase with radius as in the
observed profile. As expected (see fig.2 above), if initial number
density profile D is adopted (fig.10b), larger 
initial values of $r_a$ ($r_a>200$ kpc) are required to obtain final
velocity dispersion profiles fitting the data.

In fig. 11 we compare  the
theoretical $\sigma_R^2/\sigma_T^2$ profile obtained in the simulations leading
to final GCMF properties and velocity dispersion profile consistent
with observations and the $\sigma_R^2/\sigma_T^2$ profile
derived from our fits in fig.2; fig.11 shows that the results from our
simulations are  consistent with the constraints on the
anisotropy imposed by observational data.

Finally, in fig.12 we compare the observed surface density profile
with the final surface density profile from our simulation with
initial density profile C and initial $r_a$ equal to 60 kpc. Final
surface density profiles from simulations with $r_a>30$ kpc are
essentially indistinguishable from that shown in fig.12. 
This figure shows that a 
density profile initially more peaked than the observed one can evolve,
as a result of the preferential disruption of inner clusters, toward a final
shape in good agreement with the observed one.

Fig.13 shows the radial dependence of the fraction of surviving clusters;
assuming a similar initial density profile for stars and for the GCS,
this plot clearly shows  
that disruption can produce the observed central flattening of the GCS
density profile and  explain a rising local specific frequency up to
about 20-25 kpc. 

A large fraction of the total initial GCS mass is lost because of cluster 
disruption and mass loss from surviving clusters. Although the
majority of the mass lost is from clusters in the  inner regions
of the galaxy, a full understanding of the relevance 
 of this mass to the build-up of the central
black hole would require a careful study of the orbital evolution of
the mass lost from clusters which is  
beyond the scope of this paper.
 
The dependence of the mass loss on the initial mass of
clusters is shown in Fig. 14. This figure  shows
the ratio of the final to the initial mass of surviving clusters,
$M_f/M_i$, versus the initial mass of 
clusters from one of our simulations: as expected, most of the
surviving low-mass clusters have lost more than half of their initial
mass but it is interesting to point out that this  
is also the case for a large number of clusters with high initial 
mass. 
 Although, as expected, low-mass clusters are preferentially
disrupted by dynamical processes, a significant fraction of clusters
with large initial masses are disrupted as
well; for example, for a model with initial density profile C and
$r_a=60$ kpc,  about 40    
per cent of the initial population of clusters with $\log M_i\simeq 5$
is disrupted by evaporation in 15 Gyr.

The results of the simulations discussed in this subsection show that,
starting from a two-slope power-law initial GCMF 
with parameters similar
to those currently observed in old GCSs, it is possible to produce
models satisfying all the observational constraints (the two-slope
power-law GCMF adopted is bell-shaped 
if binned in log M and the results would be similar if other
bell-shaped GCMFs with the same initial turnover were adopted). Additional
observational data, particularly on the kinematics of the M87 GCS,
would be valuable to further constrain the range of viable initial
conditions.
\section{Discussion}
 The simulations carried out in this investigation to explore the
range of viable initial conditions for the M87 GCS and the role of
evolutionary processes have shown
that models with a power-law initial GCMF require a very strong
initial radial anisotropy to avoid the development of a steep mean
mass radial gradient inconsistent with the roughly constant $\lmf$
radial profile observed. On the other hand, adopting the
strong radial anisotropy required to obtain a flat $\lmf$ radial
profile leads to  final kinematical properties inconsistent with the
kinematical data available, which are best-fit by models with a much
weaker radial anisotropy. 
We have explored the role of several factors which could have affected
this conclusion but we found it to be very robust. In particular we
have shown that a steep $\lmf$ radial gradient 
in models with kinematical properties matching the observed
kinematical data develops in a few gigayears and even if the current
M87 potential has not been in place for one Hubble time, a
$\lmf$ radial gradient inconsistent with observational data is
still expected. We have also considered the effect of a radial
variation of the mass-to-light ratio, but the $\lmf$ radial gradient
obtained from the observed radial profile of the mean luminosity is
still much smaller than that 
produced in theoretical models.
In our simulations we started with an initial GCMF independent on the
galactocentric distance; although it would be possible to assume a
GCMF with an ad hoc initial radial variation to counteract the gradient
produced by evolutionary processes, there is neither theoretical nor
observational support for such a choice.

As already shown in a number of previous theoretical
investigations, a power-law initial GCMF  can be  
turned by the evolutionary processes considered in our study into a
bell-shaped (in $\log M$) GCMF  resembling the GCMF of old
GCSs. However, if the evolution is  
driven only by disruption and evolution due to two-body 
relaxation, interaction with the tidal field of the host galaxy and
dynamical friction, none of the models with a power-law initial GCMF
we have studied could match both 
the observed kinematical properties and the observed $\lmf$ radial profile.

Our study shows that in models starting with a bell-shaped
initial GCMF with parameters similar to those observed,  evolutionary
processes lead to final properties satisfying all the observational
constraints ($\lmf$ radial profile, spatial distribution and
kinematical data). If the initial GCMF is indeed a power-law, as both
observational studies of young cluster systems and theoretical studies
of cluster formation suggest, the details of the evolution of an
initial power-law  GCMF into a bell-shaped GCMF are still to be
clarified.

We are currently exploring this issue by
focussing our attention on 
the effects of clusters disruption (see e.g. Chernoff \& Weinberg 1990,
Fukushige \& Heggie 1995) caused by the early mass loss 
due to the evolution of massive stars (Vesperini \& Zepf 2003). 
In particular we are studying to what extent early
disruption of clusters can alter the initial GCMF and how the evolution
of the initial GCMF depends on the distribution of cluster
concentrations and on the correlation of cluster concentrations with
other cluster properties.   

\section{Conclusions}
In this paper we have carried out a number of simulations of the
M87 GCS evolution. We have considered a number of
different initial conditions and determined which can lead, driven by
the effects of internal relaxation, interaction with the tidal field
of the host galaxy and dynamical friction, to final
properties consistent with the observed GCS properties (radial profile of the
GCS mean mass, GCS spatial distribution and GCS 
kinematical properties).

We have adopted two different initial GCMF and focussed our attention
only on the functional forms and the parameters indicated by
observational studies of young and old cluster systems: a power-law similar to
that observed in young cluster systems  in merging galaxies and a
two-slope power-law with parameters similar to that currently
observed in old cluster systems. The initial velocity distribution
considered is such 
that the ratio of the radial to the tangential
velocity dispersion increases with the galactocentric distance as in
the  Osipkov-Merritt models
($\sigma_R^2/\sigma_T^2=1+r^2/r_a^2$); a wide range of  values for
the anisotropy radius have been considered ($1<r_a<250$ kpc).
NFW number density profiles with different values of the scale radius
have been studied (see Table 1 for a summary of the initial conditions
considered).

For a power-law initial GCMF we could not find any initial condition
leading to final properties satisfying all the observational
constraints. For the two-slope power-law initial GCMF considered here (but the
results are similar for other initial GCMF bell-shaped in $\log
M$ such as a log-normal or a t-Student distribution) we found a range
of initial conditions whose evolution lead to final properties
consistent with all the observational constraints.

In particular we find that:
\begin{enumerate}
\item for a power-law initial GCMF, only
 models with a strong initial radial anisotropy fit the flat $\lmf$
profile observed.
Specifically, only models with an initial value of
the anisotropy radius $r_a \ltorder 3$ kpc have a final mean mass radial
profile fairly matching
the observed $\lmf (r)$;
\item the small initial values of $r_a$ necessary to obtain a flat mean mass
radial profile from a power-law initial GCMF
lead to final kinematical properties characterized by
a strong radial anisotropy. This is inconsistent with observational kinematical
data which,
instead, favor models characterized by more isotropic velocity distributions.
\item For a two-slope power-law initial GCMF with parameters similar
to those observed in old GCSs, the final $\lmf$ radial
profile is approximately flat and consistent with 
observations for all the values of $r_a$ considered ($1<r_a<250$ kpc)
and for all the initial number density profiles considered.
\item For the two-slope power-law initial GCMF adopted in this study,
a number of the models 
studied also have final kinematical properties and final spatial
distributions consistent with observations and they, therefore, satisfy all the
observational constraints available. In particular for the initial
number density profile C (single NFW profile 
with initial scale radius equal to 0.1 kpc; see Table 1) values of
the anisotropy radius larger 30 kpc lead to final 
kinematical properties consistent with observations with the
best-fitting models being those with $50\ltorder r_a \ltorder 100$
kpc. For this range of values of $r_a$, the initially peaked spatial
distribution significantly flattens as a results of the preferential
disruption of inner clusters and the final surface number density is
perfectly consistent with observations.
For the initial number density profile D
(double NFW profile 
with scale radii equal to 0.1 and 560 kpc; see Table 1) more isotropic
initial velocity distributions are necessary to produce final
kinematical properties consistent with observations ($r_a\gtorder 200$
kpc).
\item Although low-mass clusters are those preferentially affected by
evolutionary processes, we have shown that for a significant fraction of
clusters with initial masses larger than $10^5 M_{\odot}$ the current mass
is less than 50 per cent of the initial mass.
\item We have explored the dependence of the fraction of disrupted
clusters on the galactocentric distance and shown that, although
clusters are disrupted more efficiently at small galactocentric
distances, disruption can be significant over a large radial range and
explain a rising local specific frequency out to large radii.
\end{enumerate}

In a future work (Vesperini \& Zepf 2003) we will
study the role of early disruption of clusters with different
initial concentrations caused by the mass loss due to the evolution
massive stars and we will explore the effect of this process on the
evolution of the GCMF.  
\section*{ACKNOWLEDGMENTS}
The authors gratefully acknowledge support from NASA via the ATP grant
NAG5-11320, the LTSA grant NAG5-11319 and  grant
AR-09208.01 from Space Telescope Science Institute, which is operated by AURA
under NASA contract NAS 5-26555. We thank D.E. McLaughlin for kindly
providing us with the data of the M87 GCS velocity dispersion profile and an
anonymous referee for useful comments.
\section*{REFERENCES}
Ashman K.M., Zepf S.E., 1998, Globular Cluster Systems, Cambridge
University Press \\
Baumgardt, H., 1998, A\&A, 330, 480\\
Bruzual, A.G.,  Charlot, S. 2000, private communications\\
Capuzzo Dolcetta, R., Tesseri, A., 1997, MNRAS, 292, 808\\ 
Chernoff D.F., Weinberg M.D., 1990, ApJ, 351, 121\\
Cohen, J., 2000, AJ, 119, 162\\
Cohen, J., Ryzhov, A., 1997, ApJ, 486, 230\\ 
Cote, P., et al., 2001, ApJ, 559, 828\\
Dinescu, D.I., Girard, T.M., van Altena, W.F., 1999, AJ, 117, 1792\\
Dubinski, J., 1998, ApJ, 502, 141\\
Fall, S.M., Rees, M.J., 1977, MNRAS, 181, 37P\\
Fall, S.M., Zhang, Q., 2001, ApJ, 561, 751\\
Fukushige T., Heggie, D.C., 1995, MNRAS, 276, 206\\
Gnedin O.Y., Ostriker J.P., 1997, ApJ, 474, 223\\
Harris, W.E., 2001, in Star Clusters, Lectures for 1998 Saas-Fee
Advanced School, L. Labhardt, B. Binggeli eds., (Springer-Verlag,
Berlin, 2001)\\   
Hernquist, L., 1993, ApJS, 86, 389\\ 
Kroupa, P., 2001, MNRAS, 322, 231\\
Kundu A., Whitmore B.C., Sparks W.B., Macchetto F.D., Zepf S.E.,
Ashman K.M., 1999, ApJ 513, 733\\
Kundu, A., Zepf, S.E., Ashman, K.M., 2002, in preparation\\
Lauer, T. R., Kormendy, J., 1986, ApJ, 303, L1\\
Maraston, C. et al., 2002, A\&A submitted\\
McLaughlin D.E. 1994, PASP, 106, 47\\
McLaughlin D.E. 1999a, ApJ, 512, L9\\
McLaughlin D.E. 1999b, AJ, 117, 2398\\
Merritt, D., 1985, AJ, 90, 1027\\
Murali C., Weinberg M.D., 1997a, MNRAS, 288, 767\\
Murali C., Weinberg M.D., 1997b, MNRAS, 291, 717\\
Navarro, J., Frenk, C., White, S., 1996, ApJ, 462, 563\\
Osipkov, L.P., 1979, PAZh, 5, 77\\
Rhode,K.L., Zepf, S.E., 2001, AJ, 121, 210\\ 
Romanowsky, A.J., Kochanek, C.S., 2001, ApJ, 553, 722\\
Secker,J., 1992, AJ, 104, 1472\\
Vesperini, E., 1997, MNRAS, 287, 915\\
Vesperini, E., 1998, MNRAS, 299, 1019\\
Vesperini, E., 2000, MNRAS, 318, 841\\
Vesperini, E., 2001, MNRAS, 322, 247\\
Vesperini, E., 2002, in Grebel E.K., Geisler D., Minniti D. eds.,
Extragalactic Star Clusters, p.664\\
Vesperini, E., Heggie, D.C., 1997, MNRAS, 289, 898\\
Vesperini, E., Zepf, S.E., 2003, ApJ, 587, L97\\
Whitmore, B.C., 1999, in Barnes J.E., Sanders D.B., eds., Proc. IAU
Symp. 186, Galaxy interactions at low and high redshift. Kluwer,
Dordrecht, p.251\\
Zepf, S.E., Ashman, K.M., English, J., Freeman, K.C., Sharples, R.M.,
1999, AJ, 118, 752\\
\begin{deluxetable}{cl}
\tablecaption{Summary of initial conditions. \label{tab1}}
\tablehead{
\colhead{id.} & \colhead{Number density profile: $n(r)\propto {r_s\over r} {1\over (1+r/r_s)^2}$}}
\startdata
A & $r_s=9.1$ kpc\\
B & $r_s=9.1$ for $r<95$  kpc; $r_s=560$ kpc for $r>95$ kpc \\
C & $r_s=0.1$ kpc\\
D & $r_s=0.1$ for $r<95$  kpc; $r_s=560$ kpc for $r>95$ kpc \\
\cutinhead{Initial GCMF: $f(M)\hbox{d}M\propto M^{-\alpha}\hbox{d} M$}
1PL &$\alpha=1.8$ for $10^4 M_{\odot}<M<10^7 M_{\odot}$\\
2PL &$\alpha=0.2$ for $10^4 M_{\odot}<M<10^{5.25}M_{\odot}$ and $\alpha=1.8$ for $10^{5.25}M_{\odot}<M<M<10^7 M_{\odot}$\\
\cutinhead{Initial anisotropy profile: $\sigma_R^2/\sigma_T^2=1+r^2/r_a^2$}
\nodata& $1<r_a<250$ kpc\\
 \enddata

\end{deluxetable}
\clearpage
\begin{figure*}
\plottwo{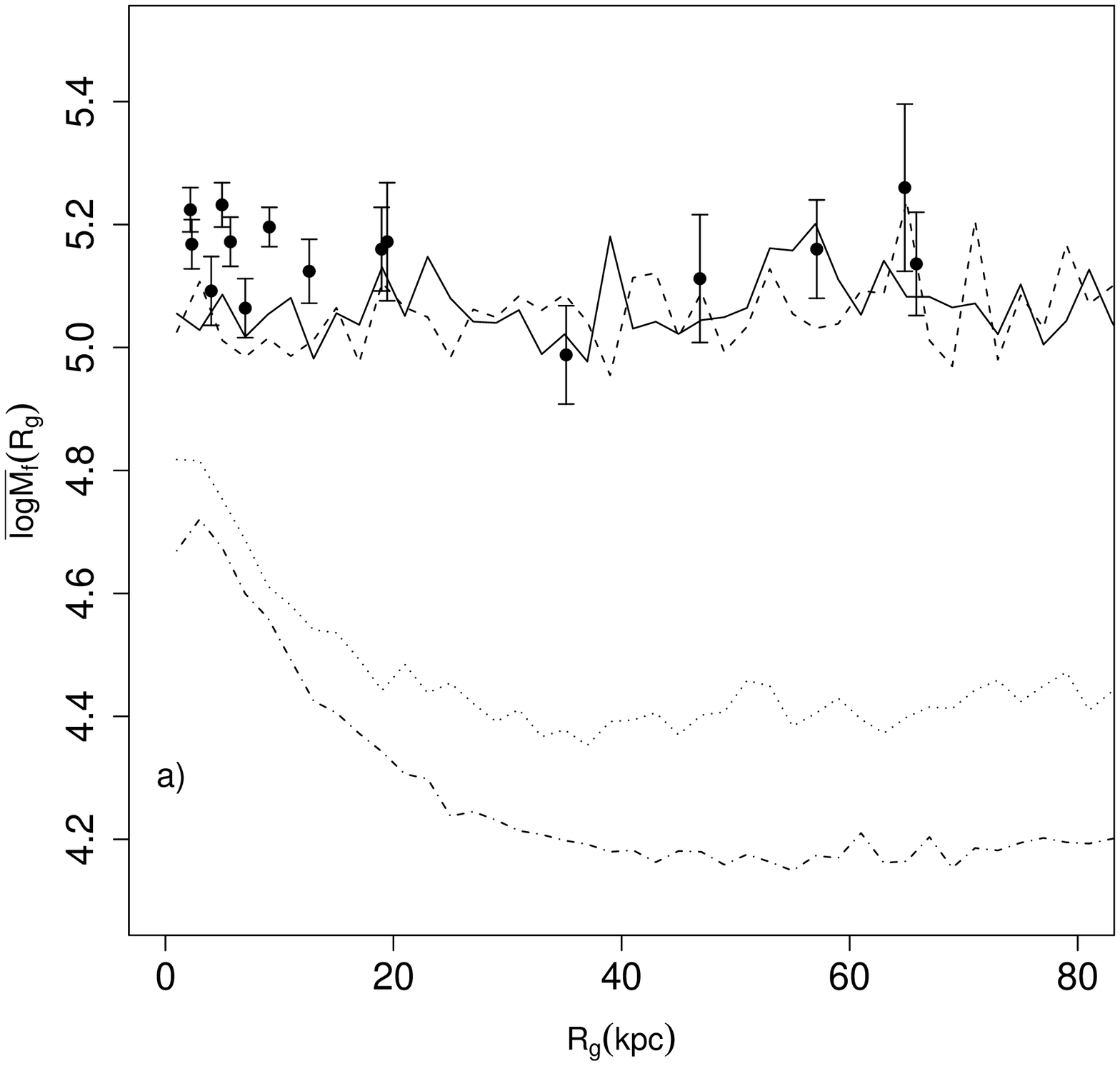}{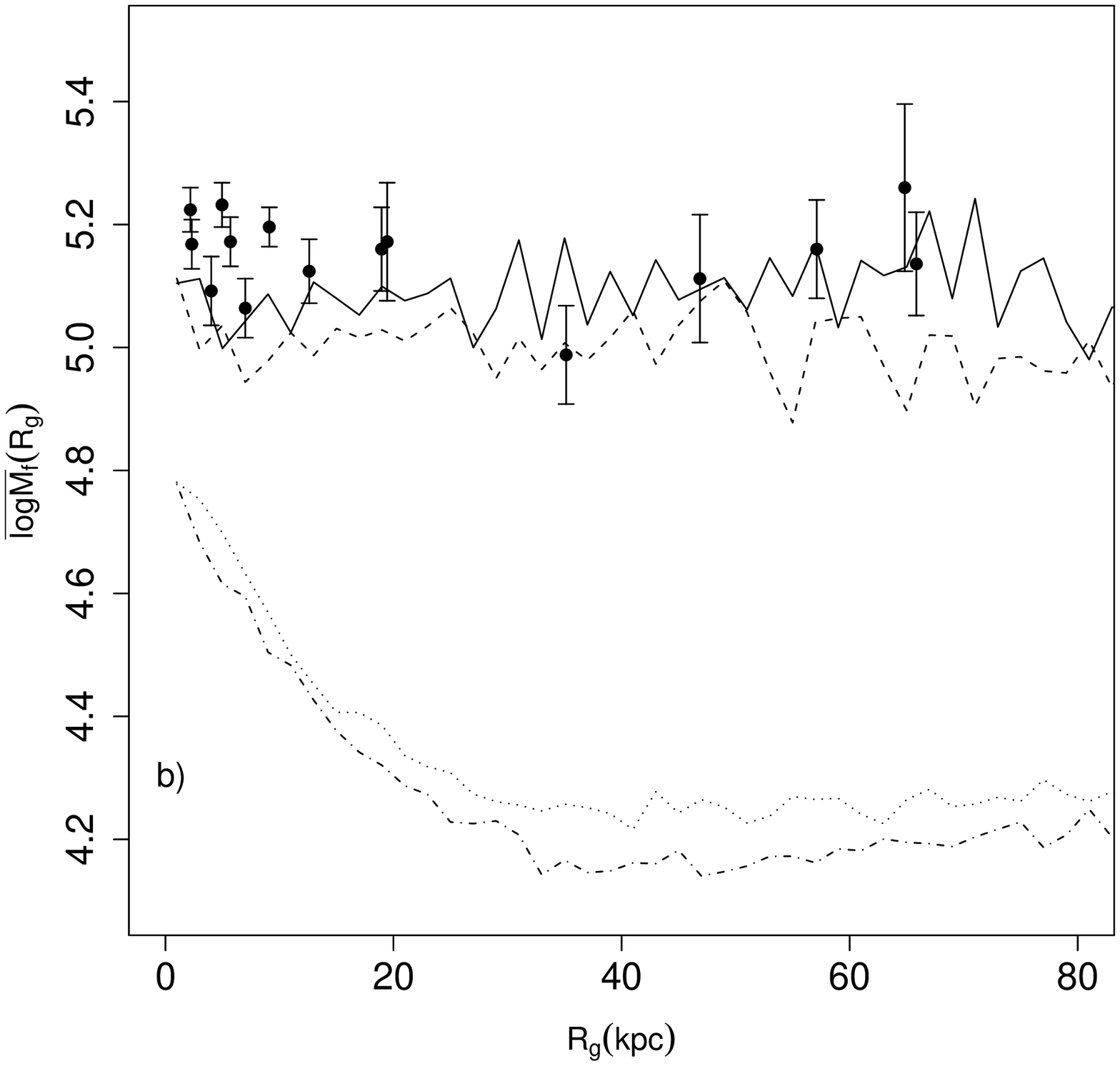}
\end{figure*}
\begin{figure*}
\plottwo{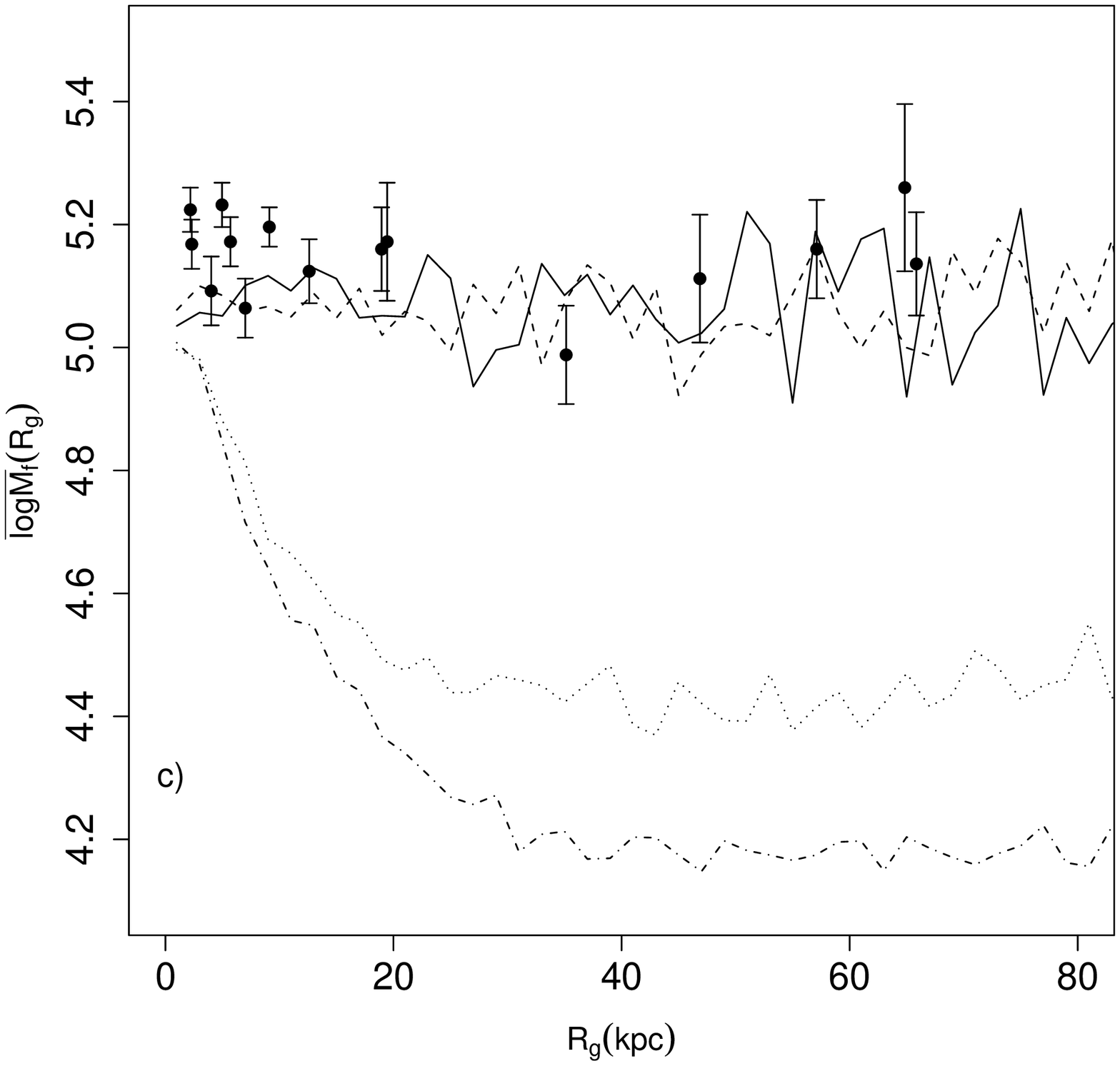}{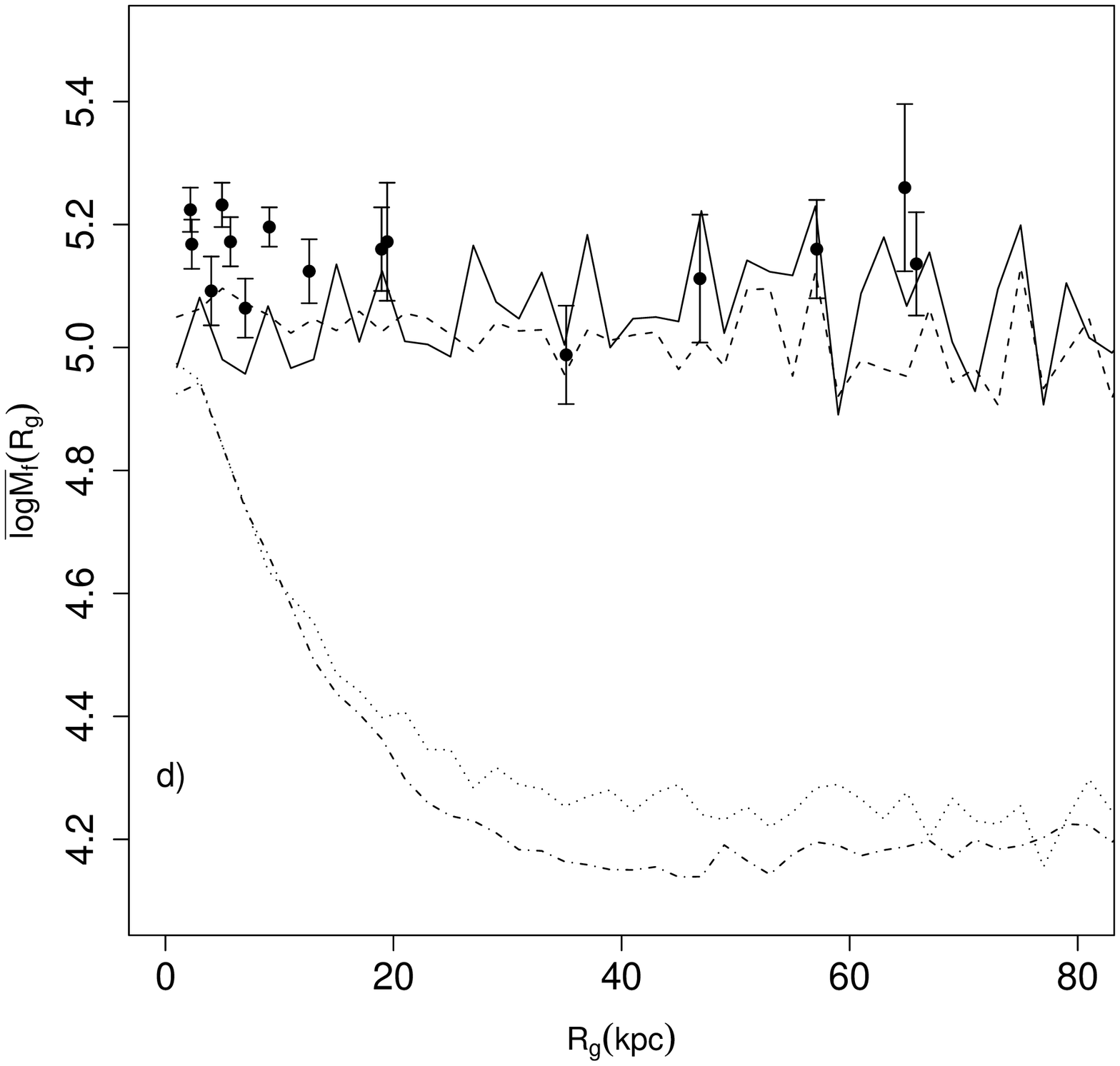}
\caption{(a)  Final GCS mean mass versus  projected galactocentric
distance for an initial anisotropy radius equal to 2 kpc (solid line), 3 kpc
(dashed line), 15 kpc (dotted line), 150 kpc (dot-dashed line) for GCS
with initial density 
profile A and a power-law initial GCMF. Dots show the mean mass gradient
as determined using Kundu et al.'s data and adopting $M/L_V=2$; (b)
Same as (a) for initial  
density profile B and $r_a$ equal to 1 kpc (solid line), 2 kpc (dashed
line), 15 kpc (dotted line), 150 (dot-dashed line); (c) same as (a)
for initial density 
profile C; (d) same as (b) for initial density profile D.}
\end{figure*}

\clearpage
\begin{figure*}
\plottwo{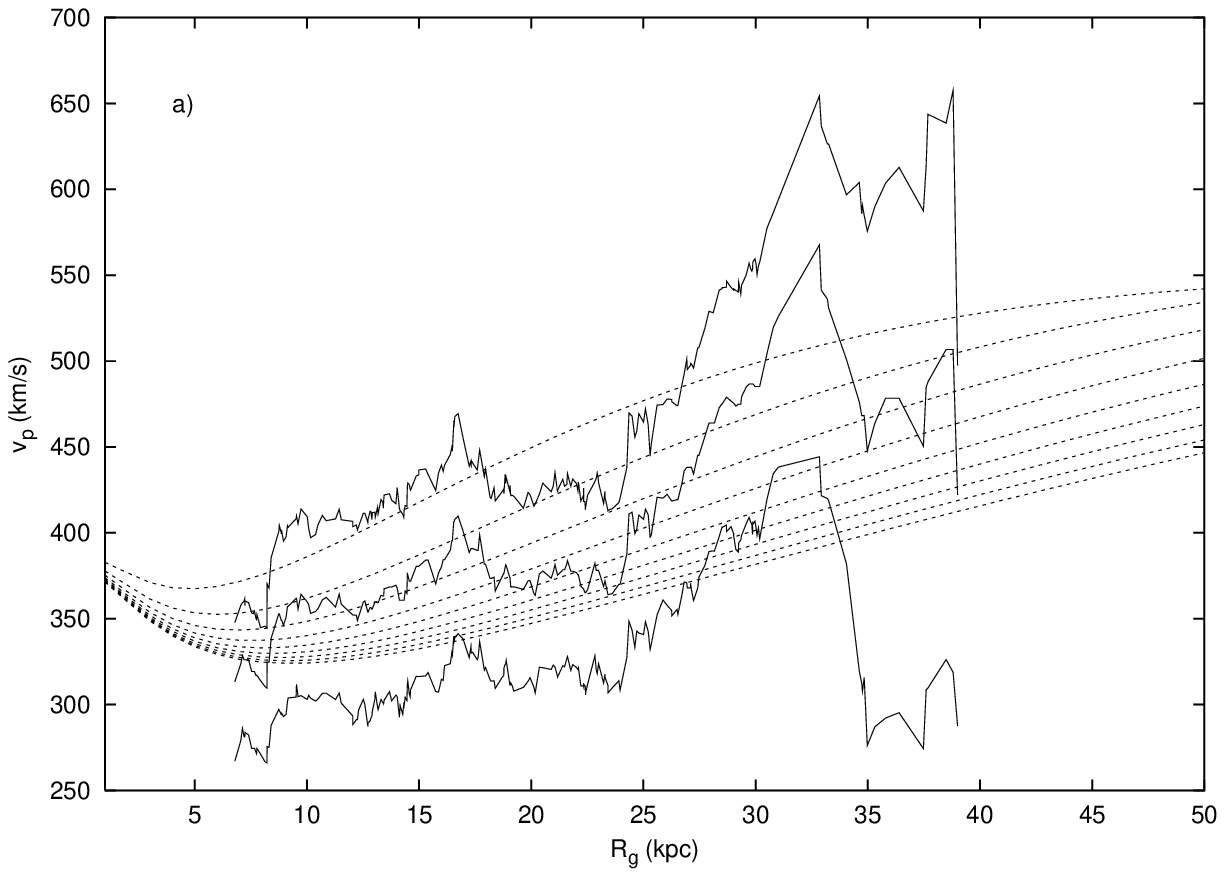}{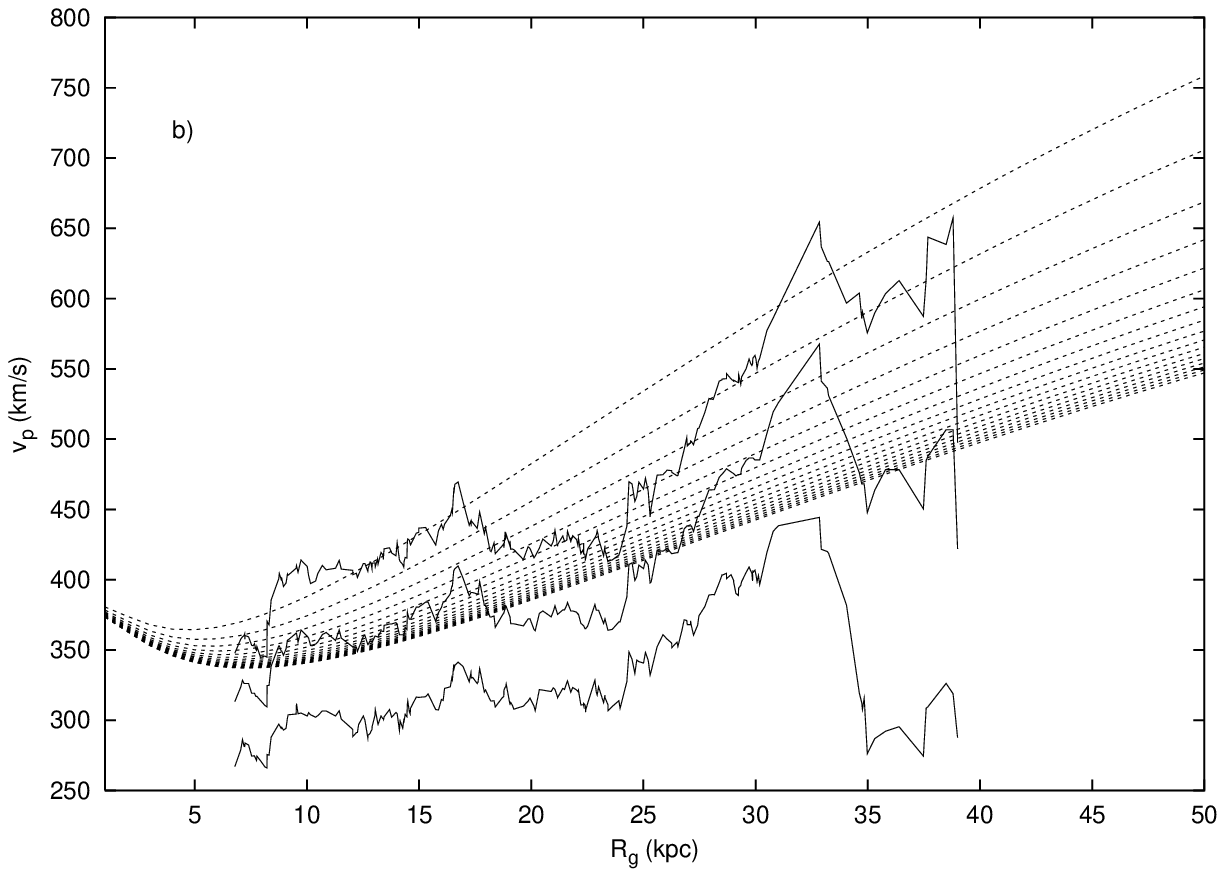}
\caption{Smoothed observed velocity dispersion radial profile of the
M87 GCS from Cote et 
al. (2001) (middle solid
line) with $90 \%$ confidence limits (upper and lower solid
lines). The dashed lines show the velocity dispersion calculated
by adopting density profile A (fig.2a) and density profile B
(fig.2b) and an Osipkov-Merrit profile for the ratio of the
radial  to tangential velocity dispersion (see equation 2 in the
text).
In fig.2a the dashed lines correspond (from top to bottom)
to values of $r_a$ ranging from 50 kpc to 210 kpc by steps of 20. 
In fig.2b the dashed lines correspond (from top to bottom)
to values of $r_a$ ranging from 200 kpc to 1000 kpc by steps of 50.} 
\end{figure*}

\begin{figure*}
\plottwo{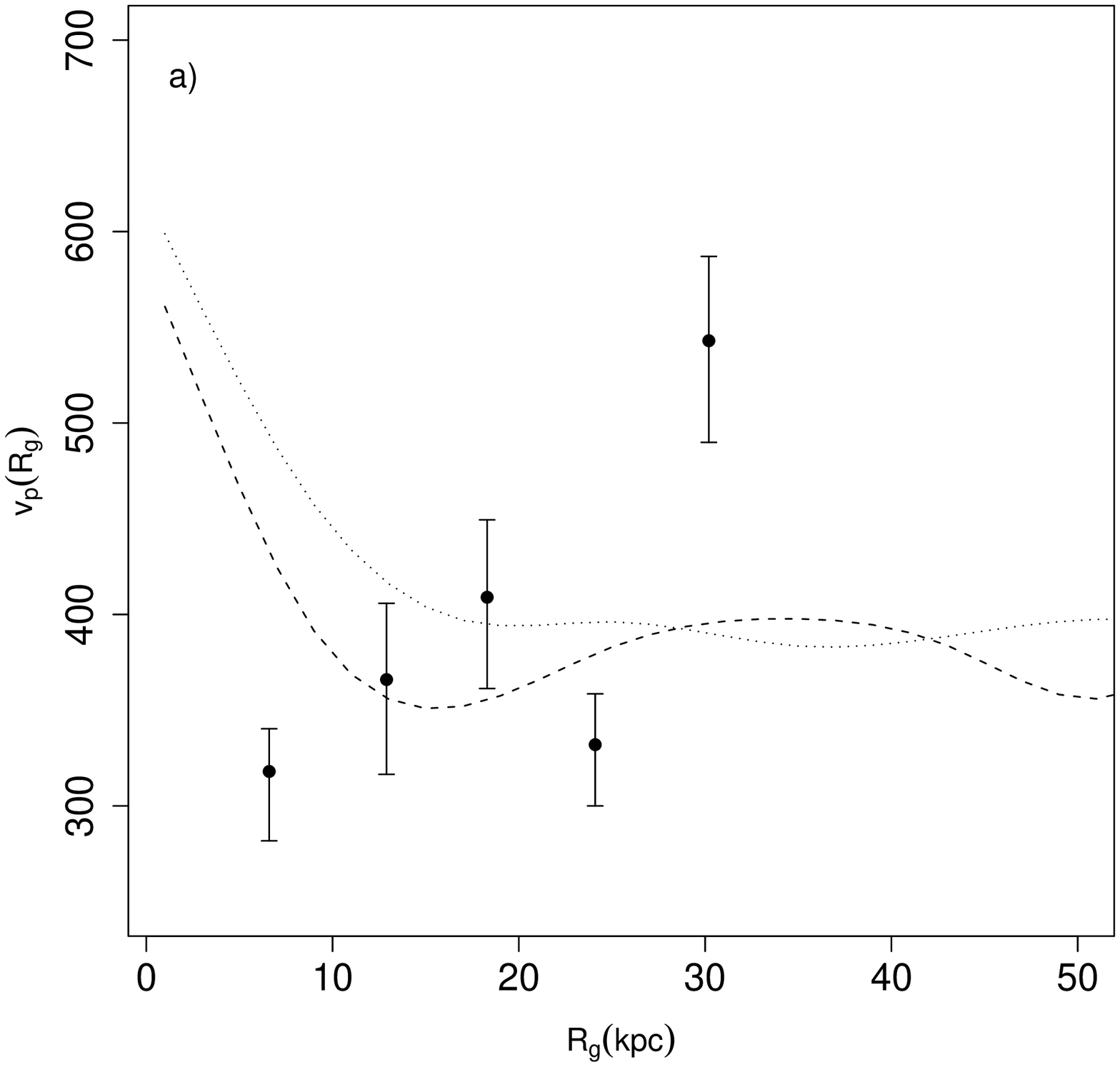}{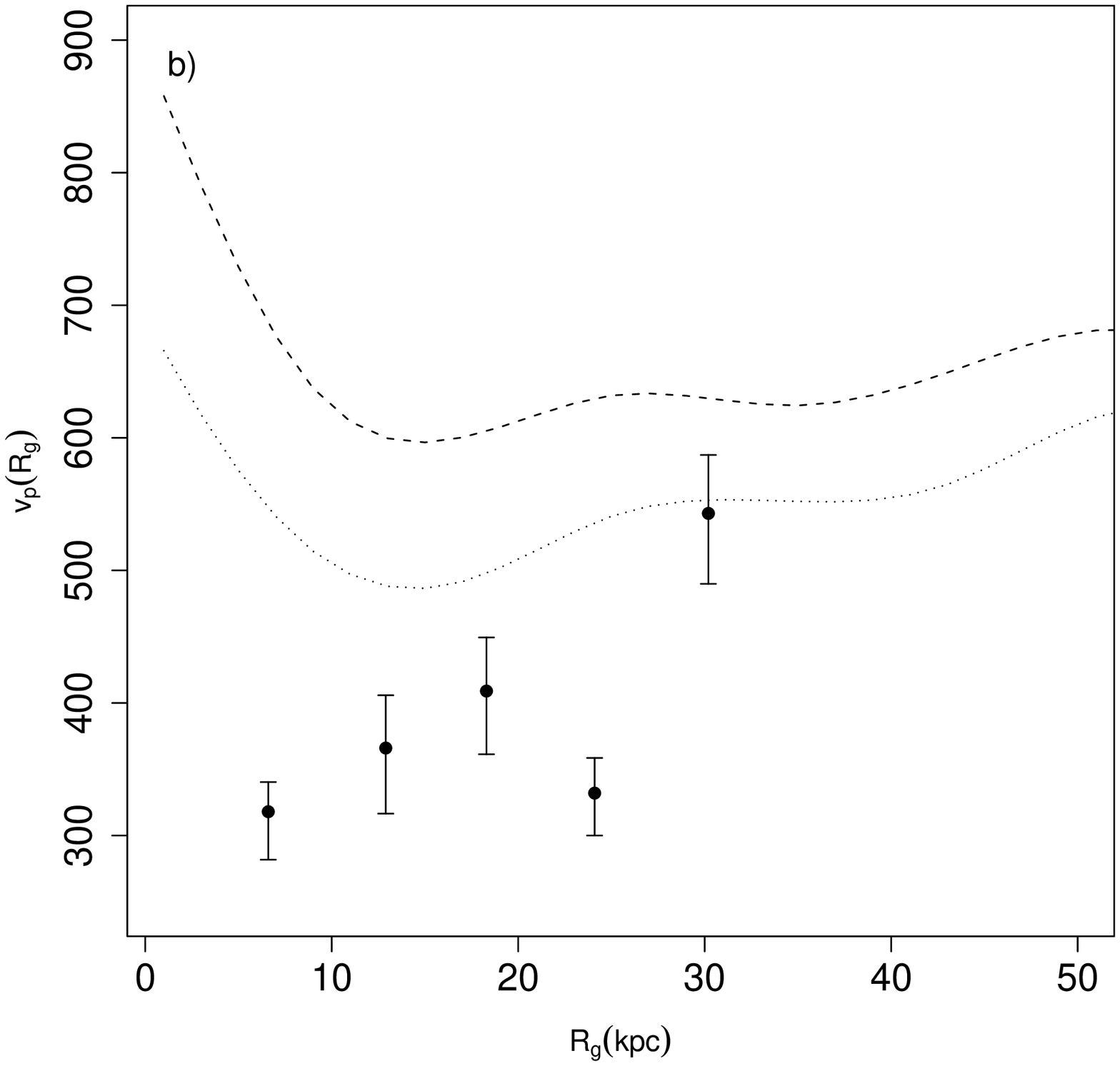}
\caption{(a) Observed velocity dispersion of the M87 GCS from Cote et
al. (2001) versus projected galactocentric distance. 
The dashed and the dotted  lines show the velocity
dispersion profiles obtained from 
our simulations of GCSs with initial density profile A and  $r_a=2$ kpc
(dashed line),  $r_a=3$ kpc (dotted line). 
(b) Same as panel (a) but for simulations of GCS with initial density
profile B and $r_a=1$ kpc
(dotted line),  $r_a=2$ kpc (dashed line).} 
\end{figure*}

\begin{figure}
\epsscale{0.9}
\plotone{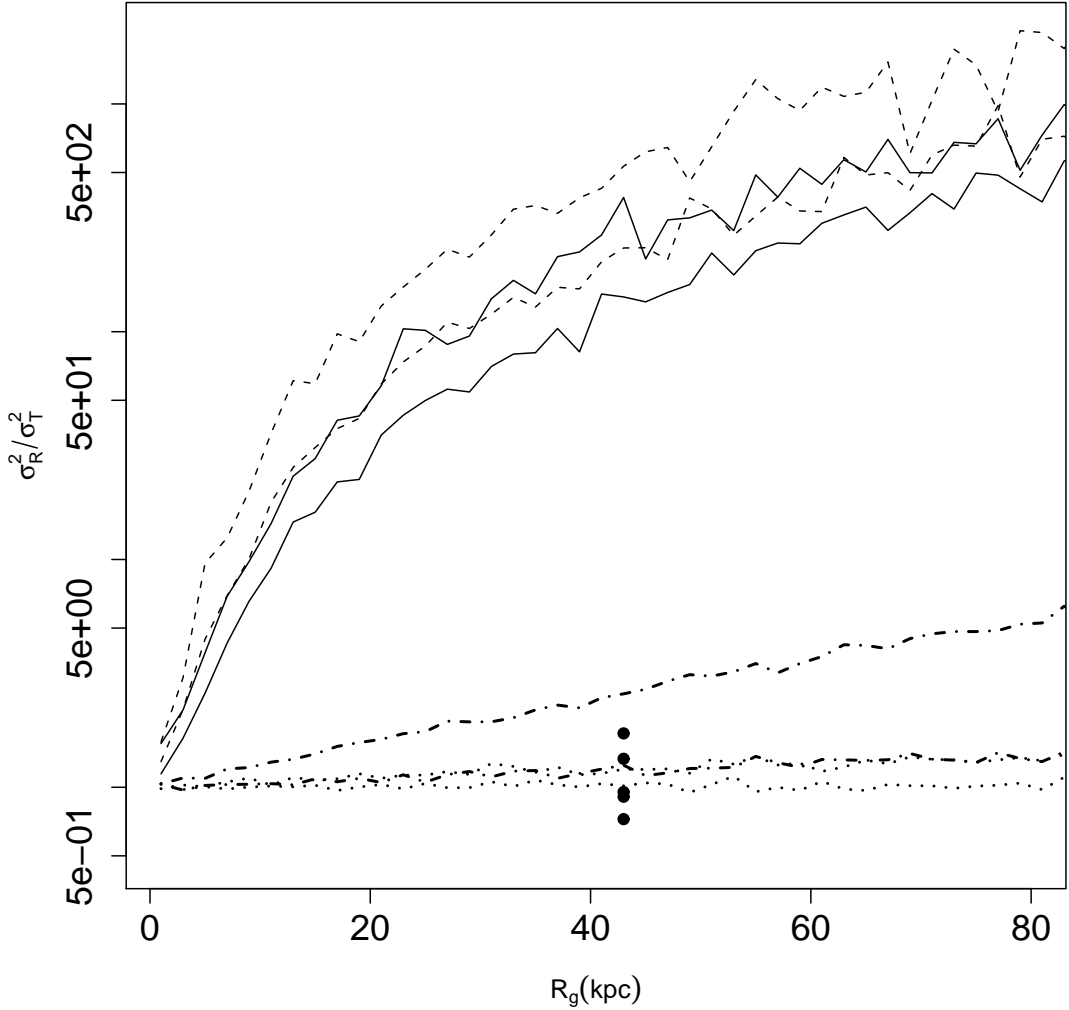}
\caption{Radial profile of the ratio of the radial to the tangential
velocity dispersion from simulations of GCSs with initial density
profile A and initial $r_a=2$ kpc (upper solid line), initial density
profile A and initial $r_a=3$ kpc (lower solid line), initial density
profile B and initial $r_a=1$ kpc (upper dashed line), initial density
profile B and initial $r_a=2$ kpc (lower dashed line).
The filled dots show the values of
$\sigma_R^2/\sigma_T^2$ at the
outermost galactocentric distance constrained by observations for the
models considered in the analysis of Romanowky \&  
Kochanek (2001). The dot-dashed lines show the radial profile of
$\sigma_R^2/\sigma_T^2$ from our fits of the observed velocity dispersion
profile assuming density profile A (see fig.2a) with $r_a=50$ kpc
(upper dot-dashed line) and $r_a=210$ kpc (lower dot-dashed line). 
The dotted lines show the radial profile of
$\sigma_R^2/\sigma_T^2$ from our fits of the observed velocity dispersion
profile assuming density profile B (see fig.2b) with $r_a=200$ kpc
(upper dotted line) and $r_a=1000$ kpc (lower dotted line). 
}
\end{figure}

\begin{figure*}
\plotone{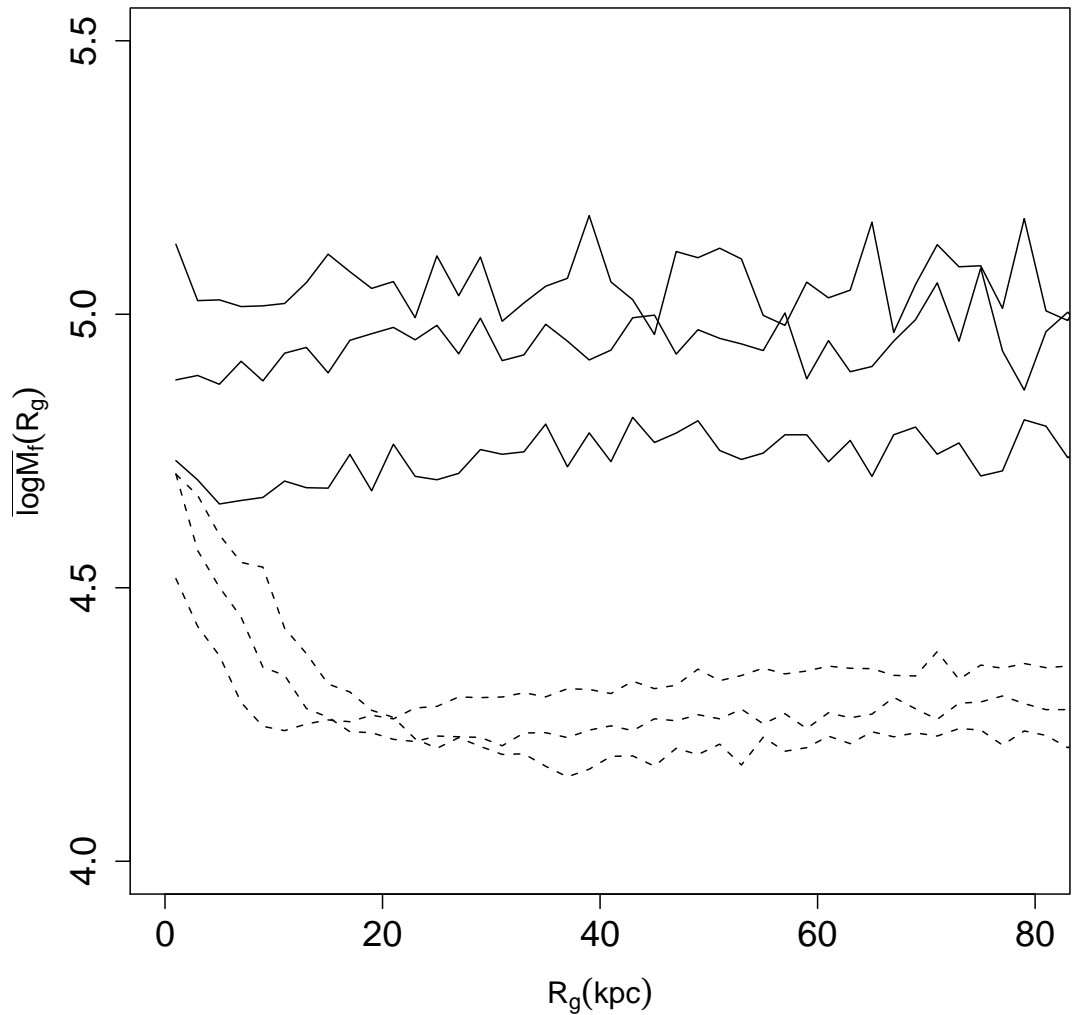}
\caption{GCS mean mass versus projected galactocentric distance for
GCS with a power-law initial GCMF, density profile A,
initial anisotropy radius equal to 2 kpc (solid lines) and 150 kpc
(dashed lines). The three lines shown for each case
correspond (from bottom to top) to the mean mass radial profile calculated  at
$t=4,~8,~12$ Gyr (for the simulation with $r_a=150$ we refer to the
order of the three lines
on the left-hand side of the plot).}
\end{figure*}

\begin{figure*}
\plottwo{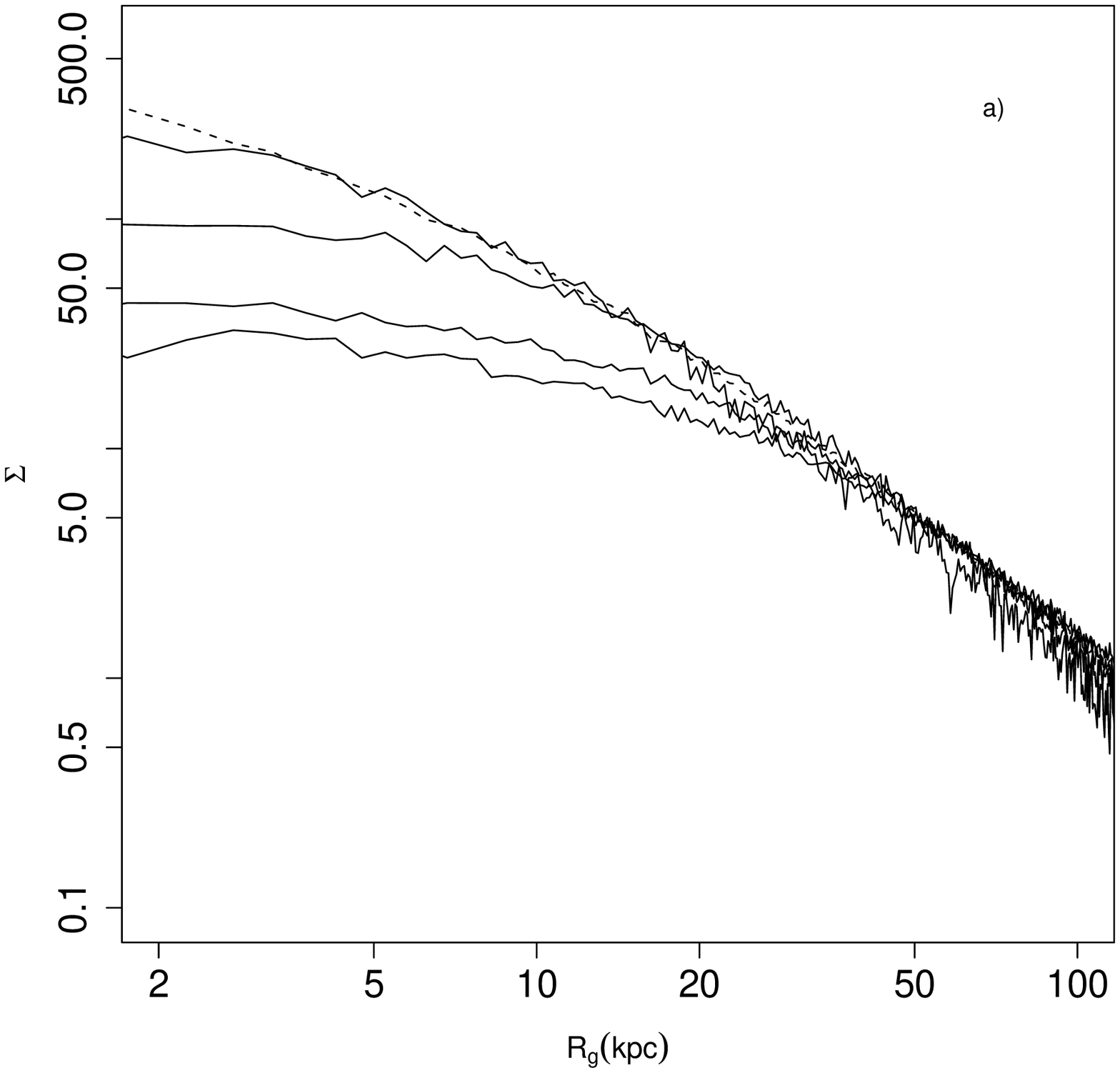}{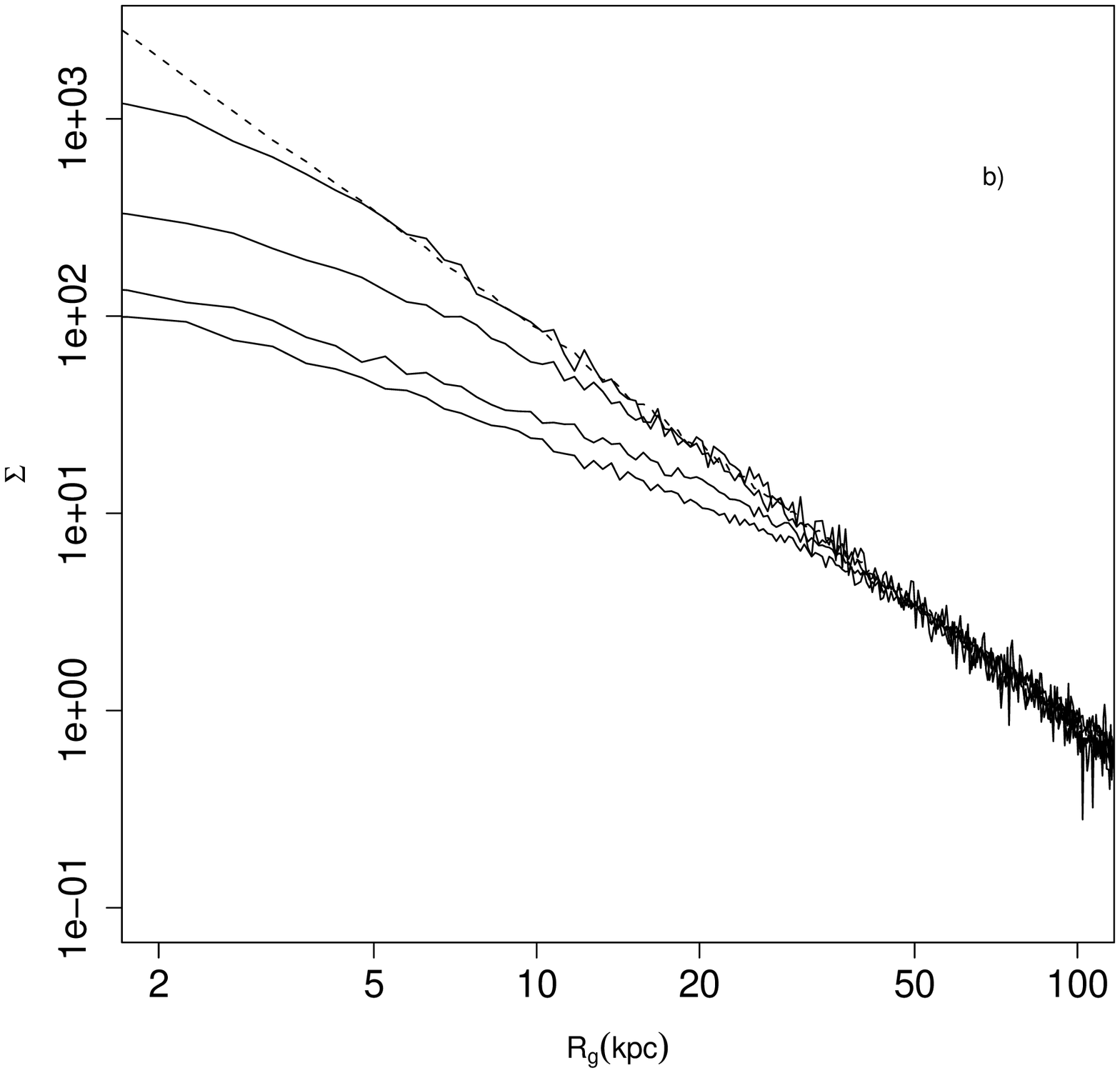}
\caption{(a) Initial (dashed line) and final (solid lines) surface
number density profile   from simulations with initial profile A and
$r_a$ (from top to bottom) equal to 1, 3, 10, 100 kpc. (b) Same as (a)
but for simulations with initial profile C. Each curve has been
arbitrarily shifted so to approximately match the initial density
profile for $R_g \gtorder 50$ kpc.}
\end{figure*}

\begin{figure}
\plotone{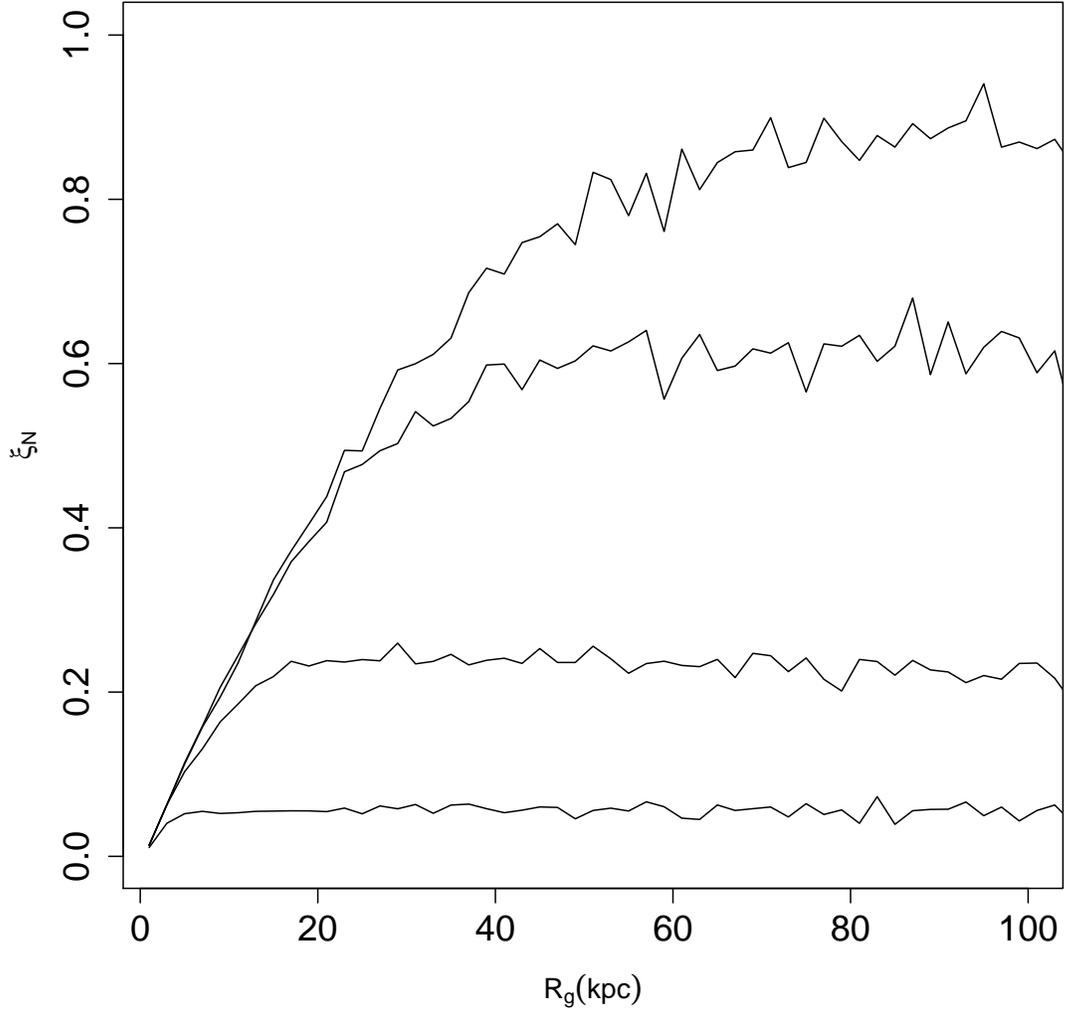}
\caption{Fraction of surviving clusters as a function of projected 
galactocentric distance for GCS with initial density profile C and
initial $r_a$ equal to (from top to bottom) 100 kpc, 30 kpc, 10 kpc,
3 kpc.} 
\end{figure}

\begin{figure*}
\plottwo{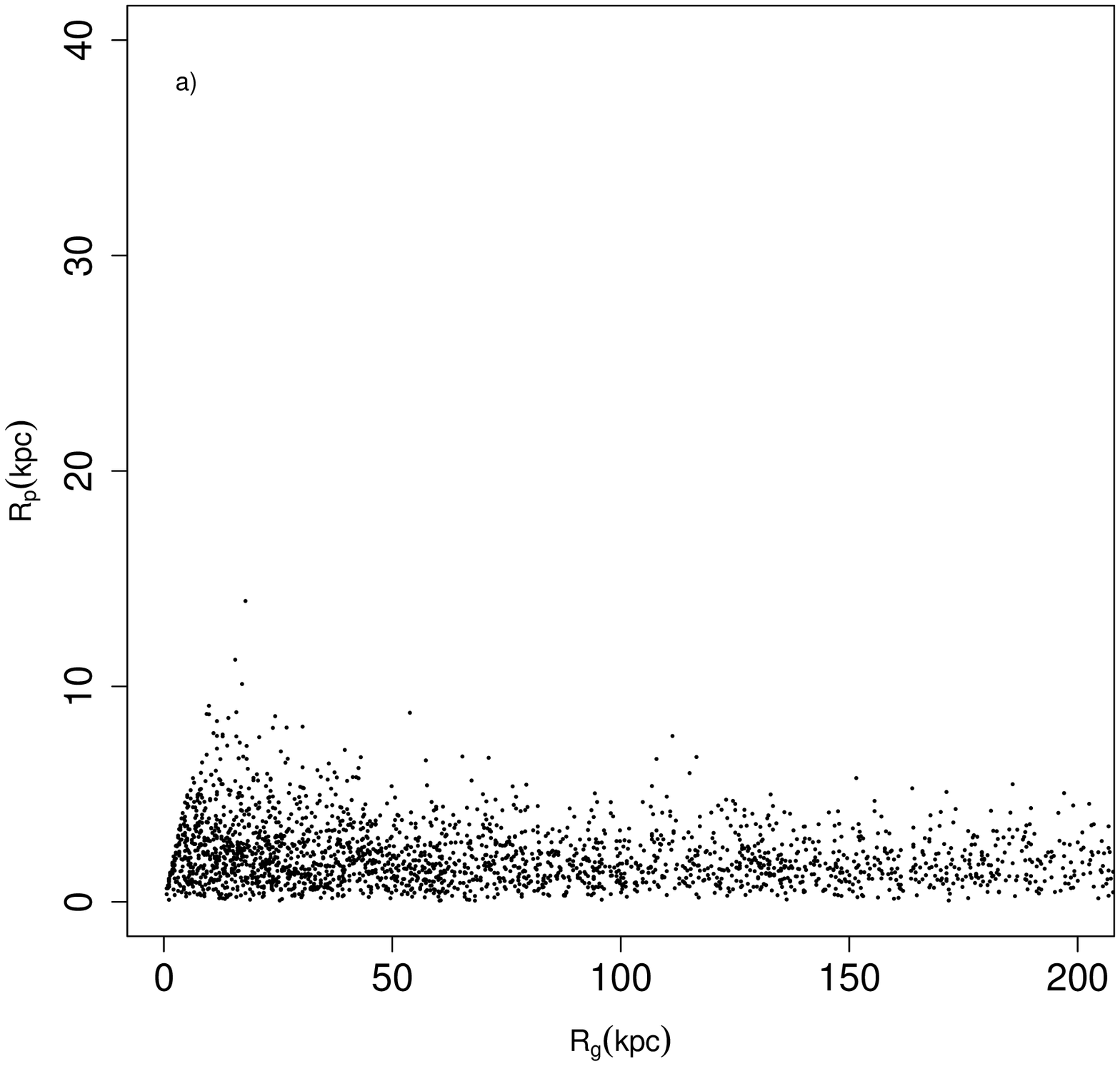}{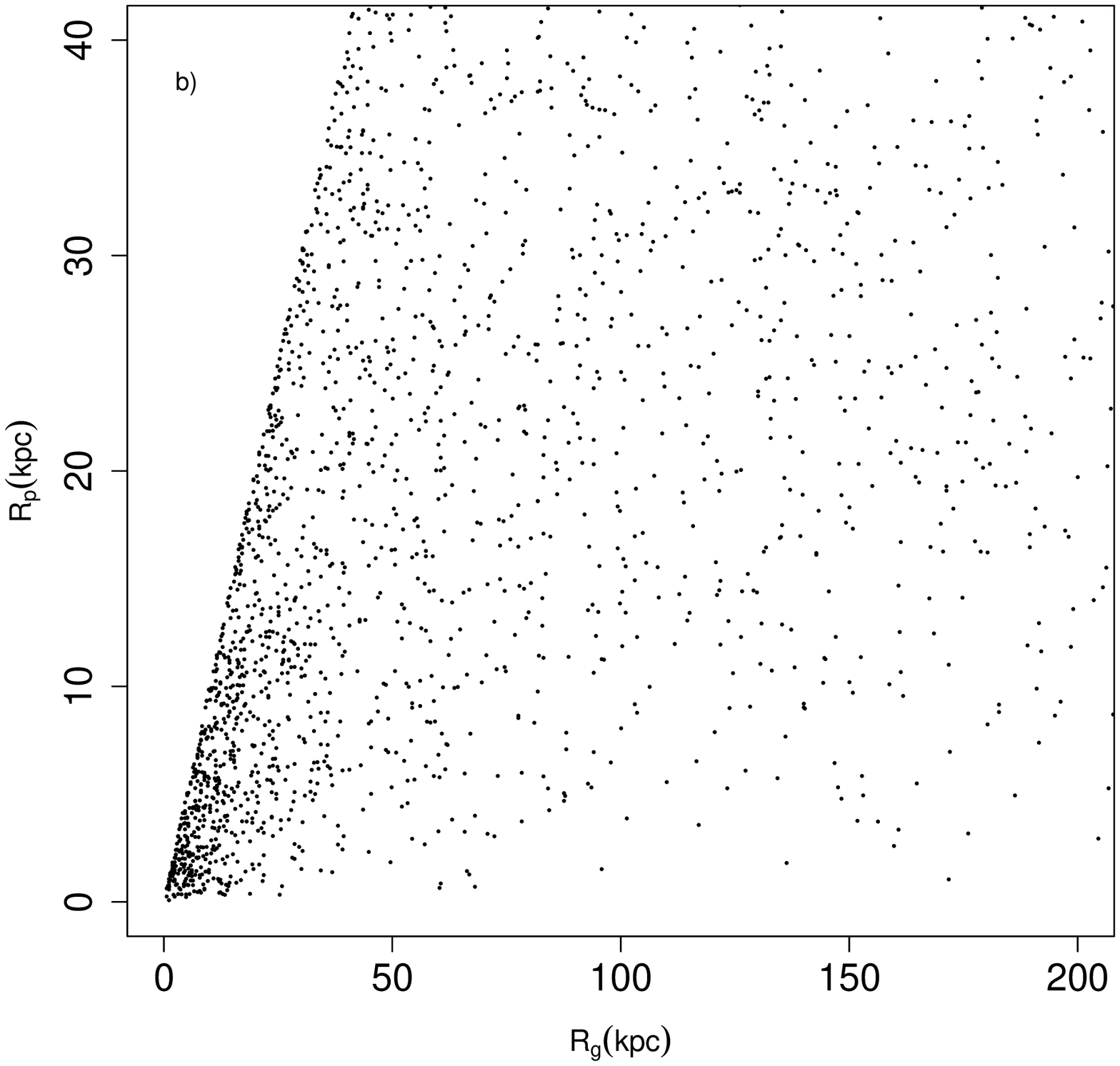}
\caption{(a) Pericenter versus current galactocentric distance for a
subsample (only 4000 out of the 400000 clusters considered in each
simulation have been plotted) of the initial population of clusters
with density profile A and $r_a=3$ kpc; (b) same as (a) but for
$r_a=50$ kpc.}
\end{figure*}

\begin{figure*}
\plottwo{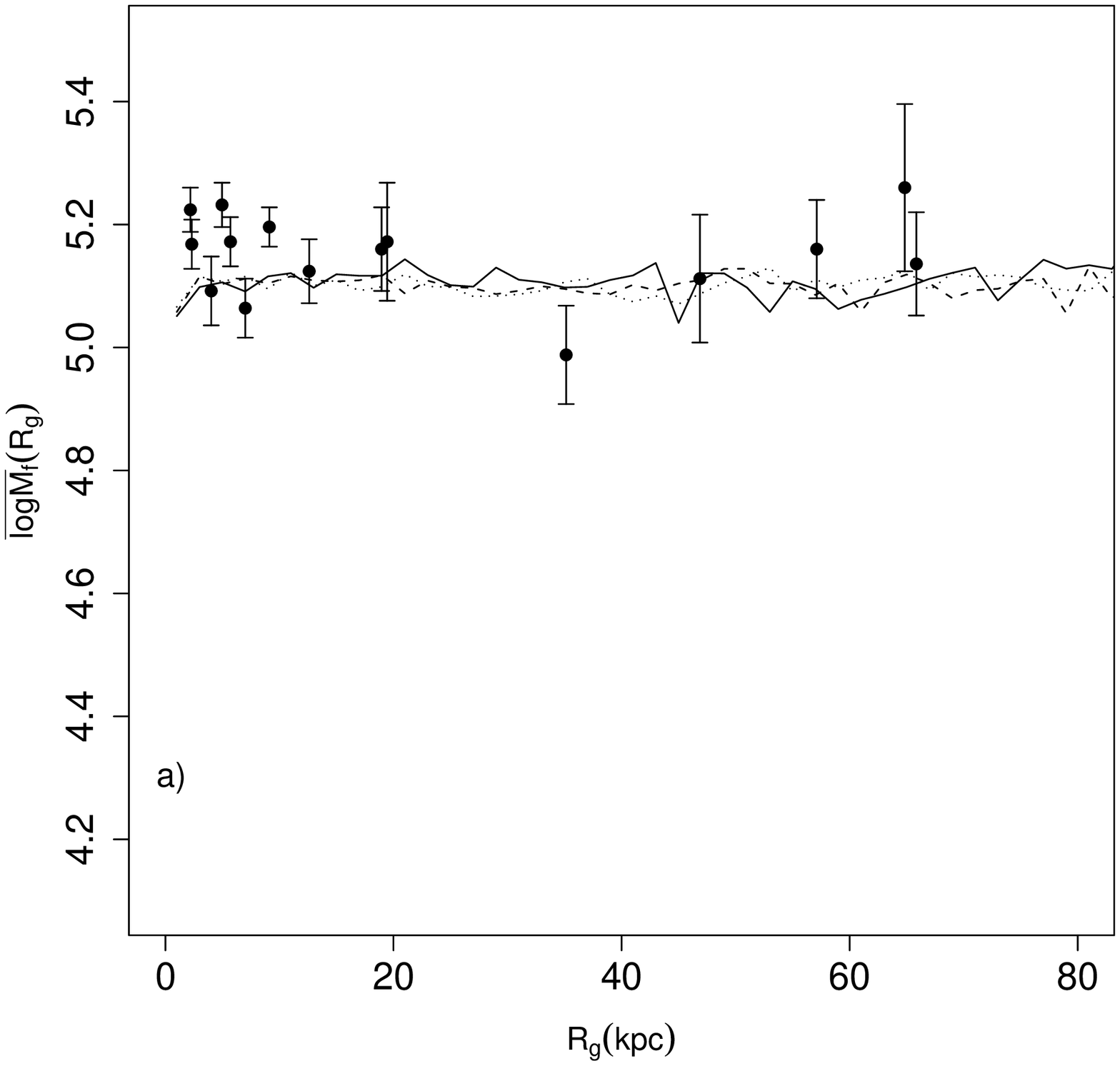}{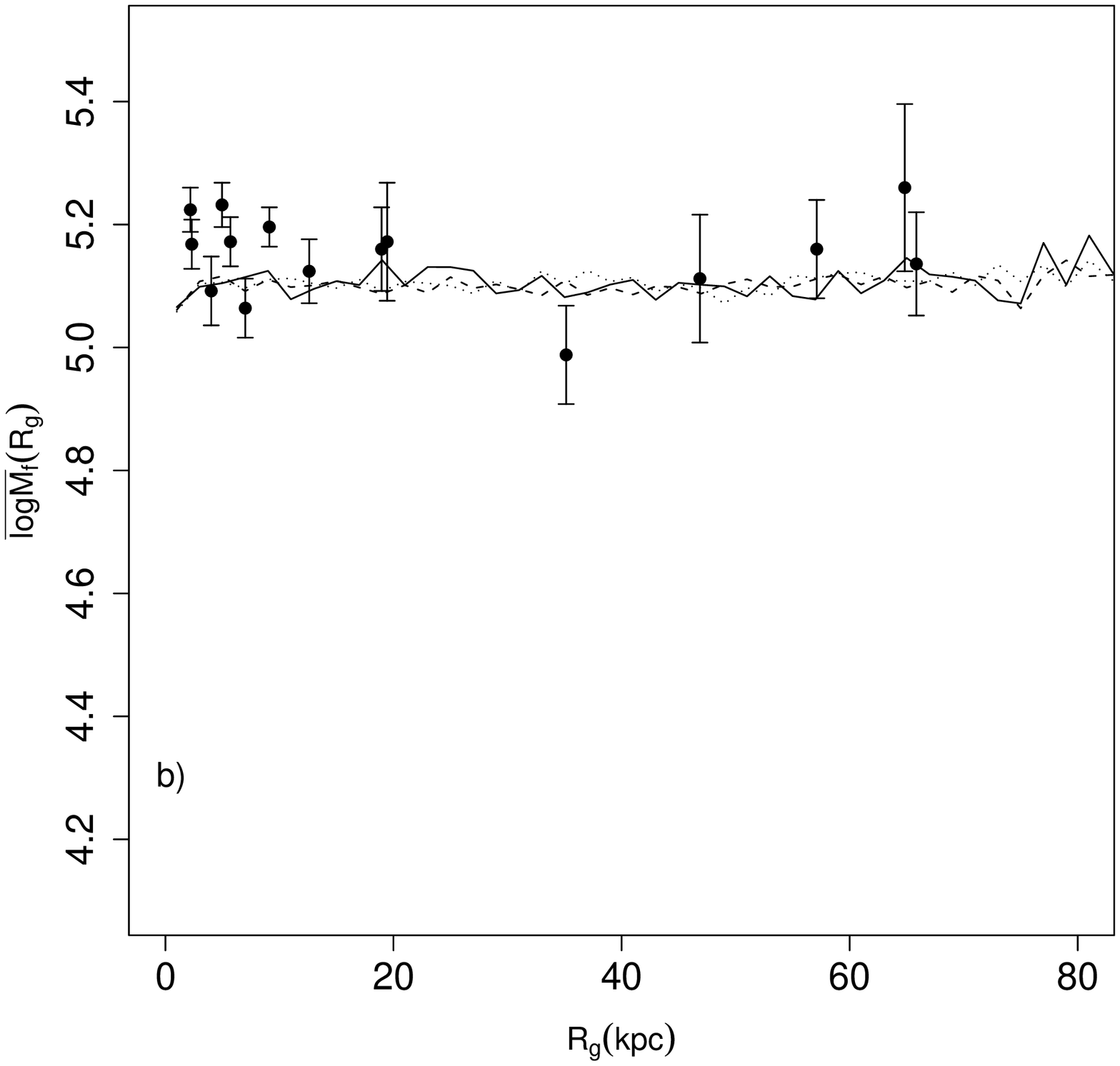}
\caption{(a)  Final GCS mean mass versus  projected galactocentric
distance for an initial anisotropy radius equal to 3 kpc
(solid line), 50 kpc 
(dashed line), 100 kpc (dotted line) initial density
profile C and a two-slope power-law initial GCMF. Dots show the mean
mass gradient 
as determined using Kundu et al.'s data and adopting $M/L_V=2$; 
(b) Same as (a) for initial
density profile D.}
\end{figure*}

\begin{figure*}
\plottwo{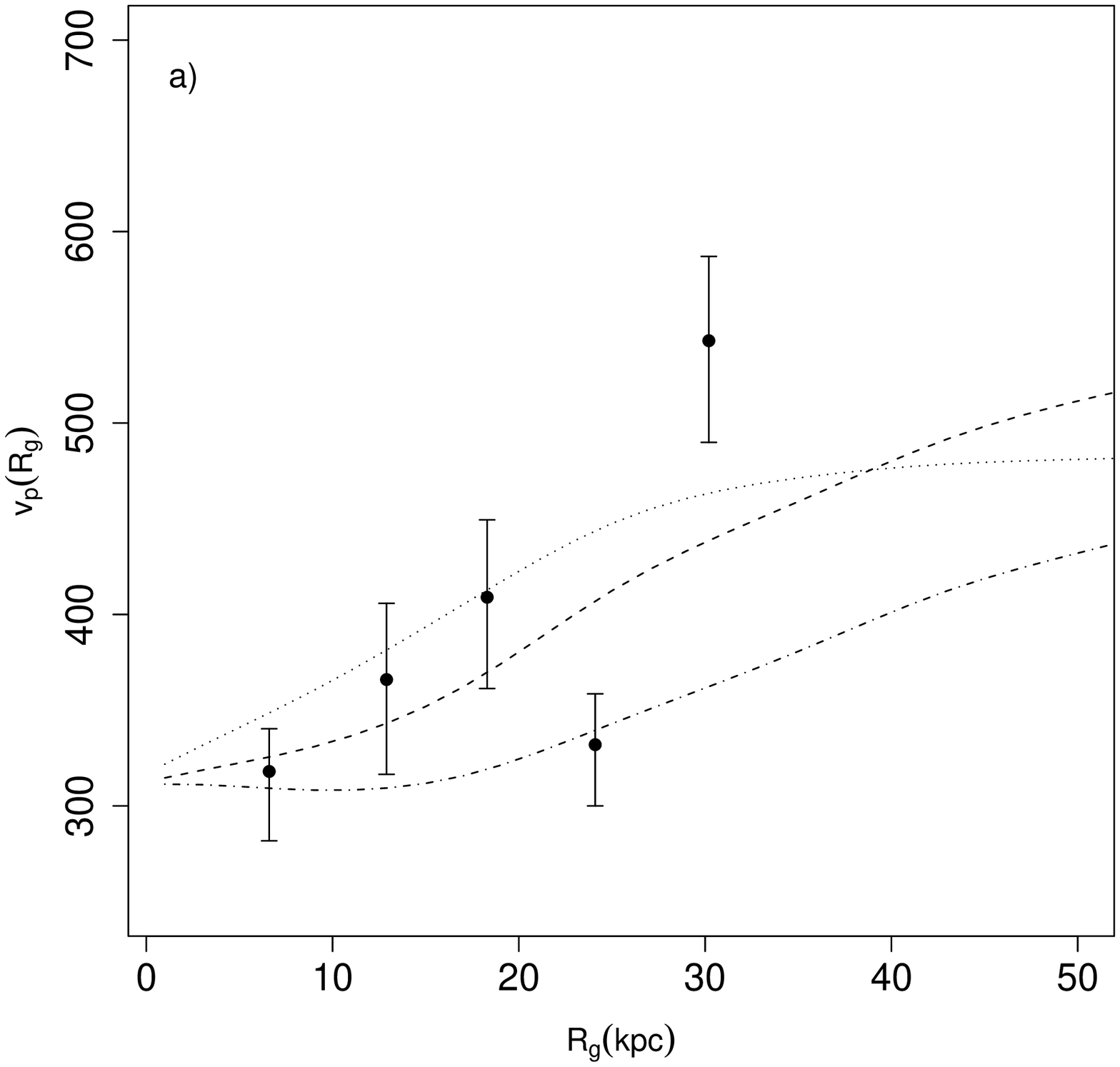}{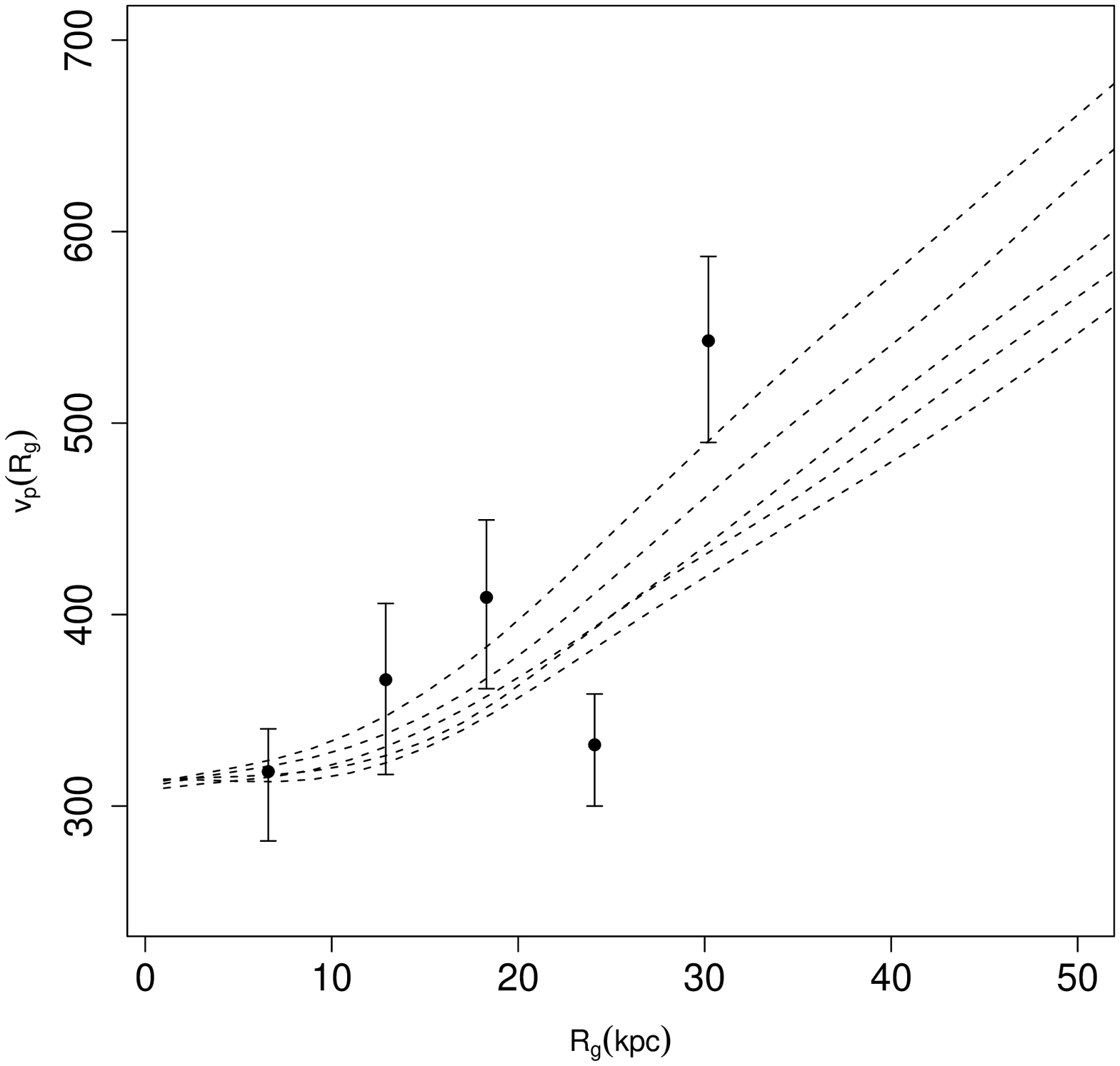}
\caption{(a) Observed velocity dispersion of the M87 GCS from Cote et
al. (2001) versus projected galactocentric distance. 
The dotted, dashed and dot-dashed  lines show the velocity
dispersion profiles obtained from 
our simulations of GCSs with initial density profile C, two-slope
power-law initial GCMF  and  $r_a=30$ kpc
(dotted line), $r_a=60$ kpc (dashed line), $r_a=200$ kpc (dot-dashed line). 
(b) Same as panel (a) but for simulations of GCS with initial density
profile D and $r_a$ equal to (from top to bottom ) 200, 250, 300, 350,
400 kpc.} 
\end{figure*}

\begin{figure}
\plottwo{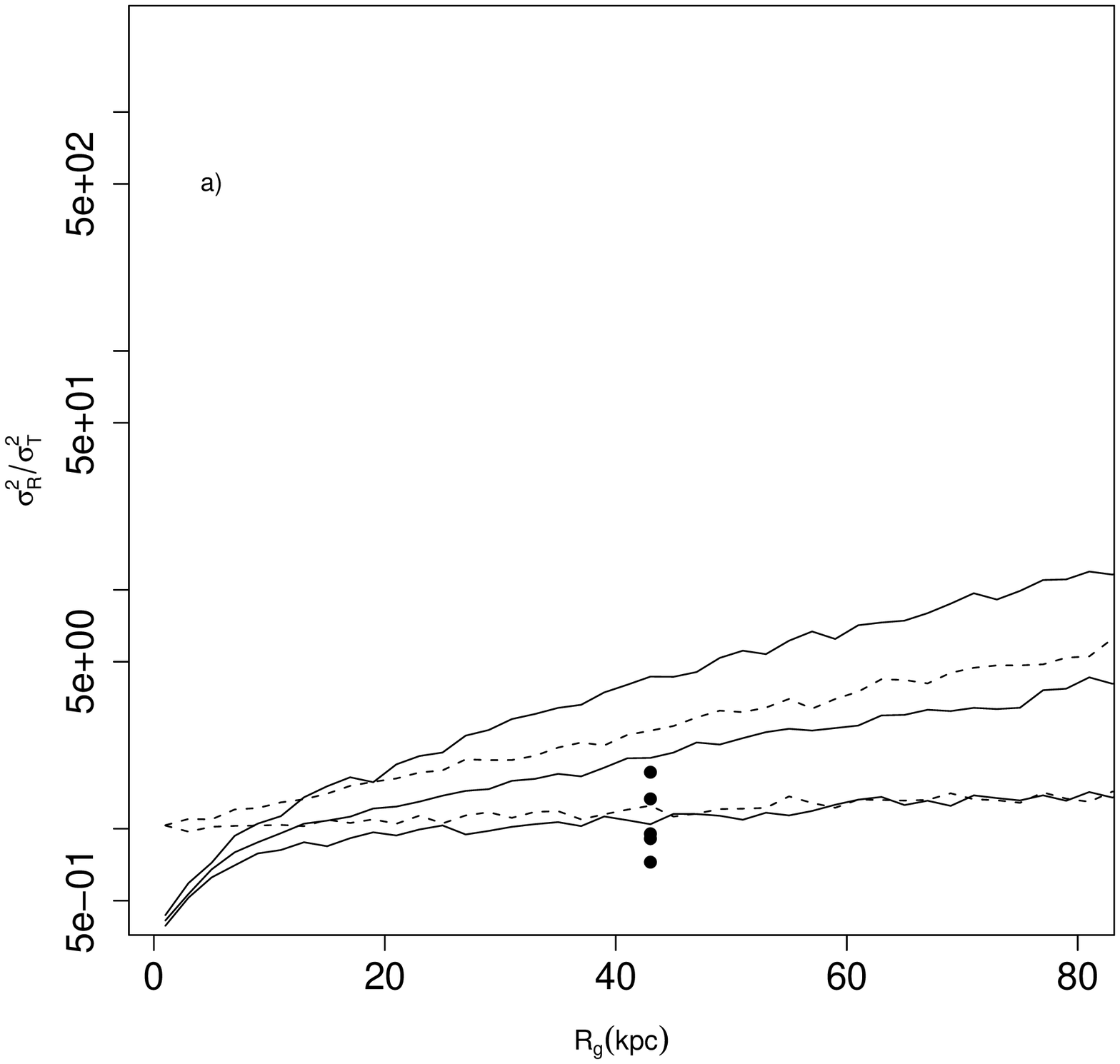}{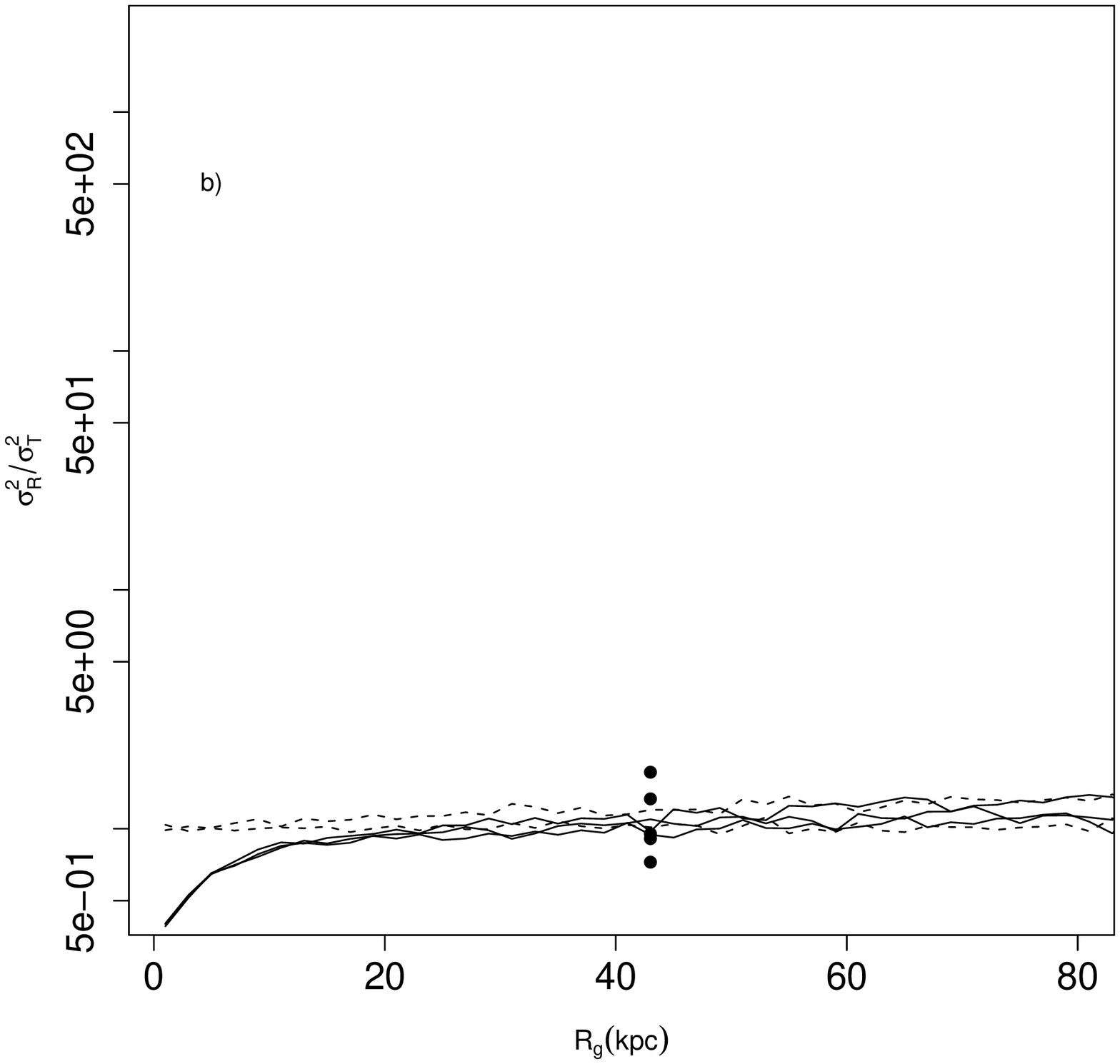}
\caption{(a) Radial profile of the ratio of the radial to the tangential
velocity dispersion from simulations of GCSs with two-slope power-law
initial GCMF, initial density
profile C  and initial $r_a$ equal to (solid lines from top to
bottom) 30, 60, 200 kpc; 
the filled dots show the values of
$\sigma_R^2/\sigma_T^2$ at the
outermost galactocentric distance for the models considered in the
analysis of Romanowky \&  
Kochanek (2001) and  the dashed lines show the radial profile of
$\sigma_R^2/\sigma_T^2$ from our fits of the observed projected velocity
profiles assuming density profile A (see fig.2a) with $r_a=50$ kpc
(upper dashed line) and $r_a=210$ kpc (lower dashed line) ; 
(b) same as (a) for initial number density
profile D  with initial $r_a$ equal to (solid lines from top to
bottom) 200, 300, 400 kpc. The dashed lines show the radial profile of
$\sigma_R^2/\sigma_T^2$ from our fits of the observed projected velocity
profiles assuming density profile B (see fig.2b) with $r_a=200$ kpc
(upper dashed line) and $r_a=1000$ kpc (lower dashed line)}
\end{figure}

\begin{figure}
\plotone{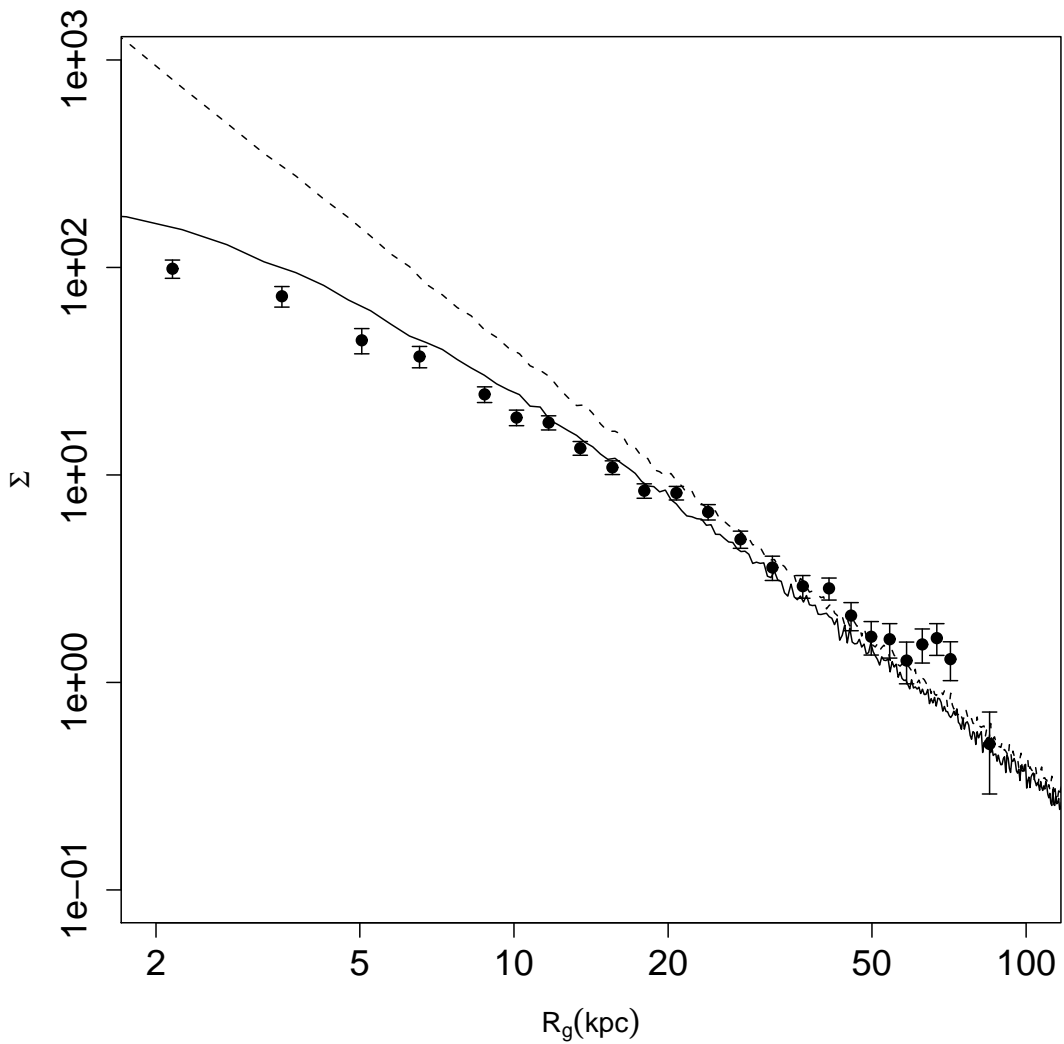}
\caption{Initial (dashed line) and final (solid line) surface number density profile from a simulation with the
two-slope power-law initial GCMF, with initial
number density profile C and $r_a=60$ kpc. Dots show the observed
surface number density profile (from McLaughlin 1999b). The final density
profile and the  
observational data have been arbitrarily shifted to approximately
match the initial density profile for $R_g \gtorder 50 $ kpc.}
\end{figure}

\begin{figure}
\plotone{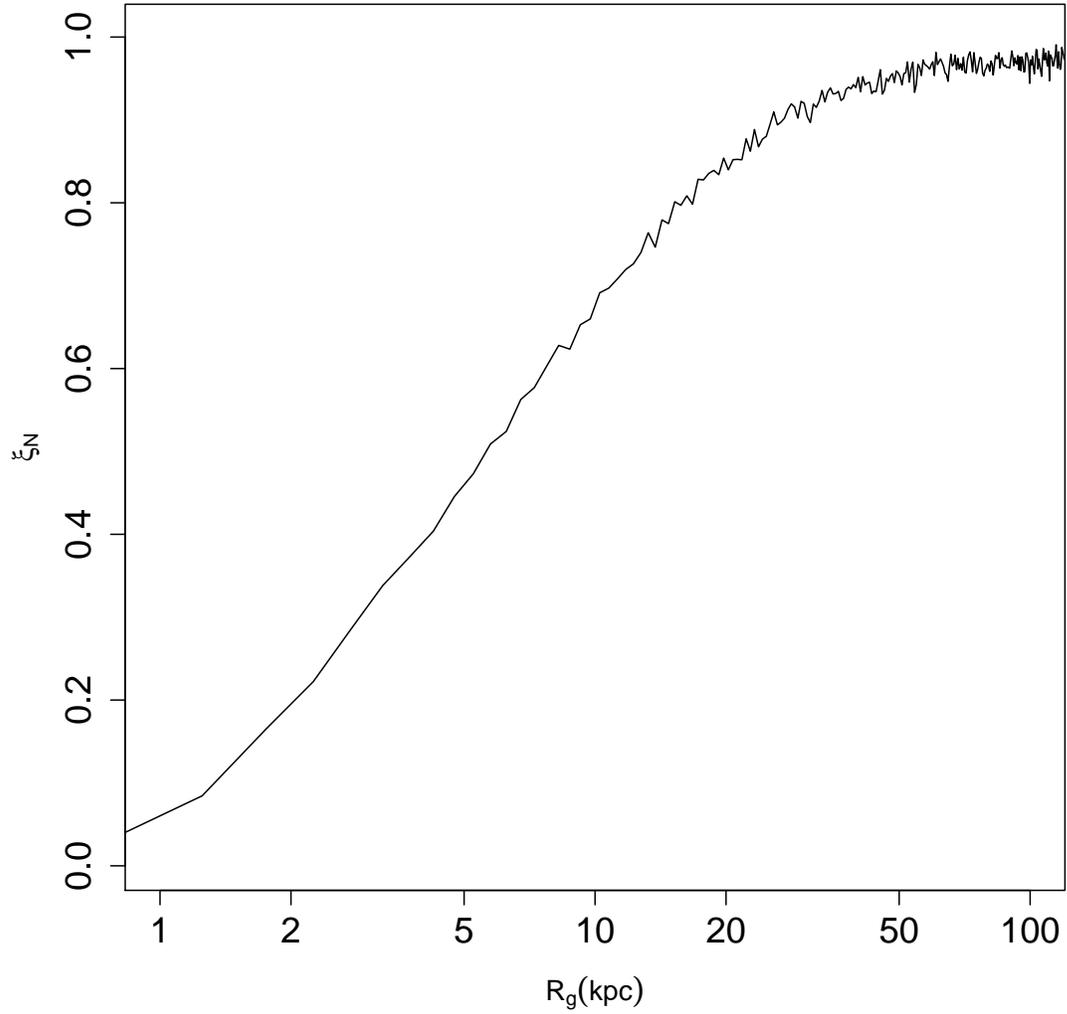}
\caption{Fraction of surviving clusters as a function of projected 
galactocentric distance for a GCS with initial density profile C,
two-slope power-law initial GCMF  and
initial $r_a$ equal to 60 kpc.} 
\end{figure}

\begin{figure}
\plotone{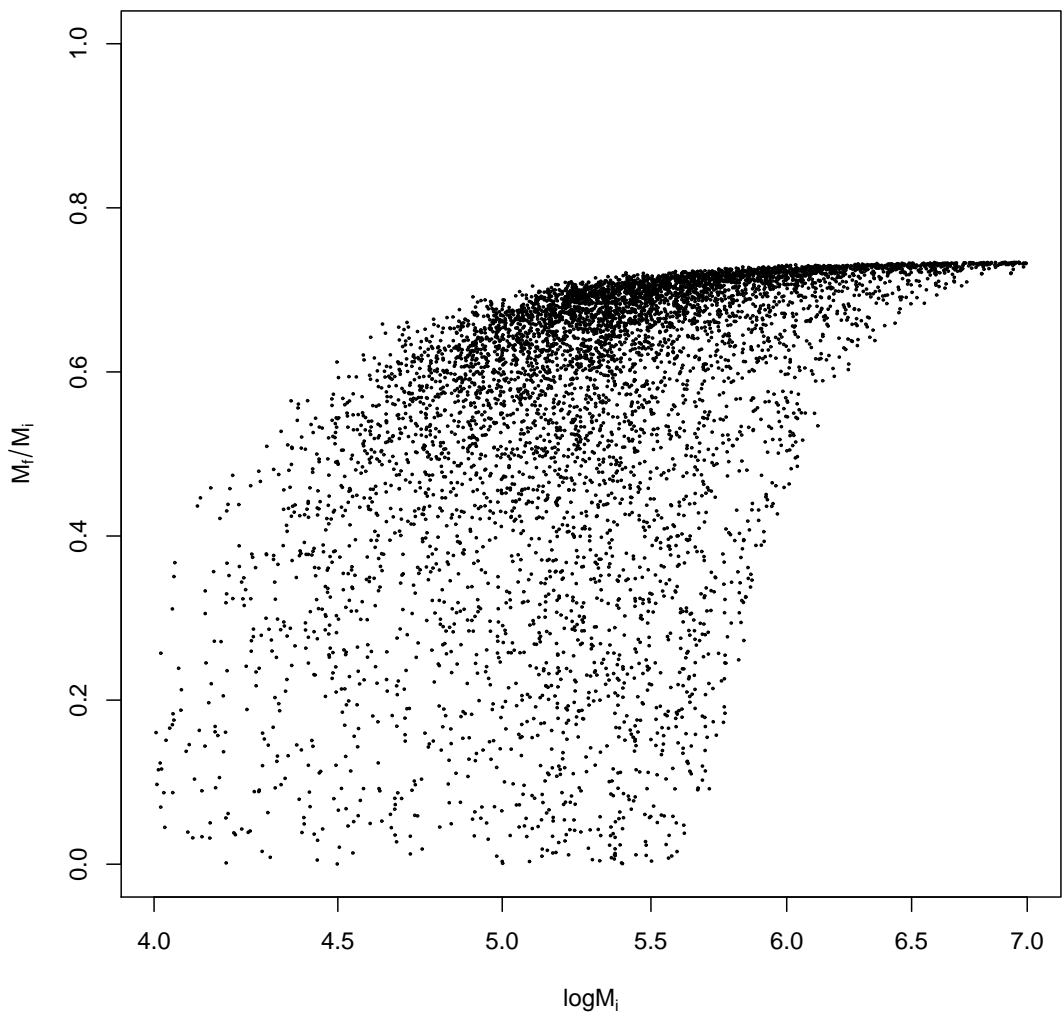}
\caption{ Final-to-initial mass ratio, $M_f/M_i$, versus logarithm
of the initial mass  for a subsample (10000 clusters)
of the population of clusters surviving for 15 Gyr from a simulation
with initial density profile C, two-slope power-law initial GCMF  and
initial $r_a$ equal to 60 kpc. Note that due to the mass loss associated
with stellar evolution all clusters lose at least about 25 per cent of their
initial mass.} 
\end{figure}

\end{document}